\documentclass[twocolumn,showpacs,nofootinbib,superscriptaddress,preprintnumbers,amsmath,amssymb,epsfig,color]{aa}

\usepackage{natbib}
\bibpunct{(}{)}{;}{a}{}{,} 

\usepackage[usenames]{color}
\usepackage{graphicx}

\usepackage{units, subfigure,amssymb,amsmath}
\usepackage[latin1]{inputenc}

\def\pbar{$\overline{p}$}

\def\vectheta{\vec{\theta}}

\begin{document}
\input epsf
\title{A Markov Chain Monte Carlo for Galactic 
cosmic-ray physics}
\subtitle{I. Method and results for the Leaky-Box Model}
\author{A. Putze\inst{1}
	\and L. Derome\inst{1}
	\and D. Maurin\inst{2}
	\and L. Perotto\inst{1,3}
	\and R. Taillet\inst{4,5}} 

\offprints{Antje Putze, {\tt putze@lpsc.in2p3.fr}}

\institute{Laboratoire de Physique Subatomique et de
	Cosmologie {\sc lpsc}, 53 avenue des Martyrs,
	Grenoble, 38026, France
	\and Laboratoire de Physique Nucl\'eaire et des Hautes
	Energies {\sc lpnhe}, Tour 33, Jussieu, Paris,
	75005, France
	\and Laboratoire de l'acc\'el\'erateur lin\'eaire {\sc lal},
	Universit\'e Paris-Sud 11, B\^atiment 200, B.P. 34, 91898 Orsay Cedex, France
	\and Laboratoire de Physique Th\'eorique {\sc lapth}, Chemin de Bellevue BP 110,
	74941 Annecy-le-Vieux, France
	\and
	Universit\'e de Savoie, Chamb\'ery, 73011, France}

\date{Received / Accepted}

\abstract
{Propagation of charged cosmic-rays in the Galaxy depends on the transport parameters,
whose number can be large depending on the propagation model under scrutiny. A
standard approach for determining these parameters is a manual scan, leading to an
inefficient and incomplete coverage of the parameter space.}
{In analyzing the data from forthcoming experiments, a more sophisticated strategy
is required. An automated statistical tool is used, which enables a full coverage
of the parameter space and provides a sound determination of the transport and
source parameters. The uncertainties in these parameters are also derived.
}
{We implement a Markov Chain Monte Carlo (MCMC), which is well suited to multi-parameter
determination. Its specificities (burn-in length, acceptance, and correlation length) are
discussed in the context of cosmic-ray physics. Its capabilities and performances are explored
in the phenomenologically well-understood Leaky-Box Model.
}
{From a technical point of view, a trial function based on binary-space partitioning
is found to be extremely efficient, allowing a simultaneous determination of up to nine parameters,
including transport and source parameters, such as slope and abundances. 
Our best-fit model includes both a low energy cut-off and reacceleration, whose values are
consistent with those found in diffusion models. A Kolmogorov spectrum for the diffusion
slope ($\delta=1/3$) is excluded.
The marginalised probability-density function for $\delta$ and $\alpha$ (the slope of the
source spectra) are $\delta\approx 0.55-0.60$ and $\alpha \approx 2.14-2.17$, depending
on the dataset used and the number of free parameters in the fit. All source-spectrum
parameters (slope and abundances) are positively correlated among themselves and with
the reacceleration strength, but are negatively correlated with the other propagation
parameters.}
{The MCMC is a practical and powerful tool for cosmic-ray physic analyses. It can be used
to confirm hypotheses concerning source spectra (e.g., whether $\alpha_i\neq \alpha_j$)
and/or determine whether different datasets are compatible.
A forthcoming study will extend our analysis to more physical diffusion models.}


\maketitle


\section{Introduction}

One issue of cosmic-ray (CR) physics is the determination of the transport
parameters in the Galaxy. This determination is based on the analysis
of the secondary-to-primary ratio (e.g., B/C, sub-Fe/Fe), for which
the dependence on the source spectra is negligible, and the ratio remains
instead mainly sensitive to the propagation processes (e.g., \citealt{2001ApJ...555..585M}
and references therein).
For almost 20 years, the determination of these parameters relied
mostly on the most constraining data, namely the HEAO-3 data, taken in
1979, which covered the $\sim 1-35$~GeV/n range \citep{1990A&A...233...96E}.

For the first time since HEAO-3, several satellite
or balloon-borne experiments (see ICRC~2007 reporter's
talk \citealt{2008arXiv0801.4534B}) have acquired higher quality 
data in the same energy range or covered a
scarcely explored range (in terms of energy, 1~TeV/n$-$PeV/n,
or in terms of nucleus): from the balloon-borne side, the ATIC collaboration has presented 
the B/C ratio at $0.5-50$~GeV/n \citep{2007arXiv0707.4415P}, and for H to Fe fluxes at 100 GeV$-$100 TeV
\citep{2006astro.ph.12377P}. At higher energy, two long-duration balloon flights
will soon provide spectra for Z=1-30 nuclei. The TRACER collaboration has
published spectra for oxygen up to iron in the GeV/n-TeV/n
range \citep{2007astro.ph..3707B,2008ApJ...678..262A}. A second long-duration flight took
place in summer 2006, during which the instrument was designed to have a wider dynamic-range
capability and to measure lighter B, C, and N elements. The CREAM experiment \citep{2004AdSpR..33.1777S}
flew a cumulative duration of 70 days in December 2004 and December 2005
(\citealt{2006cosp...36.1846S}, and preliminary results in 
\citealt{2006cosp...36.3129M} and \citealt{2006cosp...36.3231W}), and again
in December 2007. A fourth flight was scheduled for
December 2008\footnote{\tiny http://cosmicray.umd.edu/cream/cream.html}. 
Exciting data will arrive from the PAMELA satellite
\citep{2007APh....27..296P}, which was successfully launched in June
2006 \citep{2008AdSpR..42..455C}.

With this wealth of new data, it is relevant to question the method used to extract
the propagation parameters. The value of these parameters is important to many theoretical
and astrophysical questions, because they are linked, amongst others, to the transport
in turbulent magnetic fields,
sources of CRs, and $\gamma$-ray diffuse emission (see \citealt{2007ARNPS..57..285S}
for a recent review and references). It also proves to be crucial for indirect
dark-matter detection studies (e.g., \citealt{2004PhRvD..69f3501D}, and \citealt{2008PhRvD..77f3527D}).
The usage in the past has been based mostly on a manual or
semi-automated|hence partial|coverage of the parameter space (e.g.,
\citealt{1992ApJ...390...96W}, \citealt{1998ApJ...509..212S}, and \citealt{2001ApJ...547..264J}).
More complete scans were performed in \citet{2001ApJ...555..585M,2002A&A...394.1039M}, and \citet{2005JCAP...09..010L}, although in an inefficient
manner: the addition of a single new free parameter (as completed for example
in \citealt{2002A&A...394.1039M} compared to \citealt{2001ApJ...555..585M})
remains prohibitive in terms of computing time.
To remedy these shortcomings, we propose to use the Markov Chain Monte Carlo
(MCMC) algorithm, which is widely used in cosmological parameter estimates
(e.g., \citealt{2001CQGra..18.2677C}, \citealt{2002PhRvD..66j3511L}, and
\citealt{2005MNRAS.356..925D}). One goal of the paper is to confirm whether
the MCMC algorithm can provide similar benefits in CR physics.

The analysis is performed in the framework of the Leaky-Box Model (LBM),
a simple and widely used propagation model. This model contains most of
the CR phenomenology and is well adapted to a first implementation of the MCMC tool. 
In Sect.~\ref{s:key}, we highlight the appropriateness of the MCMC 
compared to other algorithms used in the field. In Sect.~\ref{s:MCMC},
the MCMC algorithm is presented. In Sect.~\ref{s:implementation}, this
algorithm is implemented in the LBM. In Sect.~\ref{s:results}, we discuss
the MCMC advantages and effectiveness in the field of CR physics,
and present results for the LBM. We present our conclusions in Sect.~\ref{s:conclusions}.
Application of the MCMC technique to a more up-to-date modelling,
such as diffusion models, is left to a forthcoming paper.


\section{Link between the MCMC, the CR data, and the model parameters}
\label{s:key}
Various models describe the propagation of CRs in the interstellar medium
\citep{1992ApJ...390...96W,1993A&A...267..372B,
1998ApJ...509..212S,2001ApJ...555..585M,2003A&A...410..189B,2006ApJ...642..882S,2008JCAP...10..018E}.
Each model is based on his own specific geometry and has its own set
of parameters, characterising the Galaxy properties.
The MCMC approach aims to study quantitatively how the existing
(or future) CR measurements can constrain these models
or, equivalently, how in such models the set of parameters
(and their uncertainties) can be inferred from the data. 

In practice, a given set of parameters in a propagation model implies, e.g.,
a given B/C ratio. The model parameters are constrained such as
to reproduce the measured ratio. The standard practice used to
be an eye inspection of the goodness of fit to the data. This
was replaced by the $\chi^2$ analysis in recent papers:
assuming the $\chi^2$ statistics is applicable to the problem at stake,
confidence intervals in these parameters can be extracted
(see App.~\ref{s:CL}).

The main drawback of this approach is the computing time required
to extend the calculation of the $\chi^2$ surface to a wider parameter
space. This is known as the curse of dimensionality, due to the
exponential increase in volume associated with adding extra dimensions
to the parameter space, while the {\em good} regions of this space
(for instance where the model fits the data) only fill a tiny volume. 
This is where the MCMC approach, based on the Bayesian statistics,
is superior to a grid approach. As in the grid approach, one end-product
of the analysis is the $\chi^2$ surface, but with a more efficient sampling
of the region of interest.
Moreover, as opposed to classical statistics, which is based
on the construction of estimators of the parameters, Bayesian statistics
assumes the unknown parameters to be random variables. As such, their
full distribution|the so-called conditional probability-density function (PDF)|given
some experimental data (and some prior density for these parameters, see below) 
can be generated.

To summarise, the MCMC algorithm provides the PDF of the model parameters,
based on selected experimental data (e.g., B/C).
The mean value and uncertainty in these parameters
are by-products of the PDF. The MCMC enables the enlargement of the
parameter space at a minimal computing time cost (although the MCMC and
Metropolis-Hastings algorithms used here
are not the most efficient one). The technicalities of the MCMC
are briefly described below. The reader is referred to \citet{1993Neal}
and \citet{2003it...book....M} for a more substantial coverage of the subject.


\section{Markov Chain Monte Carlo (MCMC)}
\label{s:MCMC}

Considering a model depending on $m$ parameters 
\begin{equation}
\vectheta\equiv \{ \theta^{(1)},\,\theta^{(2)}, \, \ldots, \,\theta^{(m)}\},
\label{eq:theta}
\end{equation}
we aim to determine the conditional PDF of the parameters given the data,
$P (\vectheta|\mathrm{data})$. This so-called 
{\em posterior} probability quantifies the change in the degree of belief one 
can have in the $m$ parameters of the model in the light of the data.
Applied to the parameter inference, Bayes theorem is
\begin{equation}
P (\vectheta|\mathrm{data}) =  \frac{P (\mathrm{data}|\vectheta)
\cdot P (\vectheta)}{P (\mathrm{data})}, 
\label{eq:bayes_theorem}
\end{equation}
where $P(\mathrm{data})$ is the data probability (the latter does not depend on the
parameters and hence, can be considered to be a normalisation factor). 
This theorem links the posterior probability to the
likelihood of the data ${\cal L(\vectheta})\equiv P (\mathrm{data}|\vectheta)$
and the so-called {\em prior} probability, $P (\vectheta)$, indicating the degree 
of belief one has {\em before} observing the data.
To extract information about a single parameter, $\theta^{(\alpha)}$, 
the posterior density is integrated over all other parameters 
$\theta^{(k \neq \alpha)}$ in a procedure called marginalisation. Finally, 
by integrating the individual posterior PDF further, we are able to determine
the expectation value, confidence level, or higher order mode of the parameter $\theta^{(\alpha)}$. 
This illustrates the technical difficulty of Bayesian parameter estimates:
determining the individual posterior PDF requires a high-dimensional 
integration of the overall posterior density. Thus, an efficient sampling
method for the posterior PDF is mandatory.
For models of more than a few parameters, regular grid-sampling approaches
are not applicable and statistical techniques are required \citep{1997sda..book.....C}.

Among these techniques, MCMC algorithms have been fully tried and
tested for Bayesian parameter inference \citep{2003it...book....M, 1993Neal}.
MCMC methods explore any {\em target} distribution given by a vector of
parameters $p (\vectheta)$, by generating a sequence of $n$ points (hereafter a
 chain)
\begin{equation}
\{\vec{\theta}_i\}_{i=1, \ldots, n}\equiv \{ \vectheta_1,\,\vectheta_2, \, \ldots, \,\vectheta_n\}.
\end{equation} 
Each $\vectheta_i$ is a vector of $m$ components [as defined in
Eq.~(\ref{eq:theta})]. In addition, the chain is Markovian in the sense that
 the distribution of $\vectheta_{n+1}$ is influenced entirely
by the value of $\vectheta_n$. MCMC algorithms are developed so that the time
spent by the Markov chain in a region of the parameter space is proportional to 
the target PDF value in this region. Hence, from
such a chain, one can obtain an independent sampling of the PDF.
The target PDF as well as all marginalised PDF are estimated by counting the
number of samples within the related region of parameter space.

Below, we provide a brief introduction to an MCMC using
the Metropolis-Hastings algorithm (see \citealt{1993Neal} and
\citealt[chapter 29]{2003it...book....M} for further details and references).

        \subsection{The algorithm}

The prescription that we use to generate the Markov chains from the unknown target 
distribution is the so-called Metropolis-Hastings algorithm. The Markov chain increases by 
jumping from the current point in the parameter space $\vec\theta_i$ to the following 
$\vec\theta_{i+1}$. As said before, the PDF of the new point only depends on the 
current point, i.e. $\mathcal{T}(\vectheta_{i+1}|{\vectheta_1, \ldots, \vectheta_i}) = 
\mathcal{T}(\vectheta_{i+1}|\vectheta_i)$. This quantity defines the transition 
probability for state $\vectheta_{i+1}$ from the state $\vectheta_i$.
The Metropolis-Hastings algorithm specifies $\mathcal{T}$
to ensure that the stationary distribution of the chain asymptotically
tends to the target PDF one wishes to sample from. 

At each step $i$ (corresponding to a
state $\vectheta_{i}$), a trial state $\vectheta_{\mathrm{trial}}$ is
generated from a {\em proposal} density $q(\vectheta_{\mathrm{trial}}| \vectheta_i)$.
This proposal density is chosen so that samples can be easily generated
(e.g., a Gaussian distribution centred on the current state).
The state $\vectheta_{\mathrm{trial}}$ is accepted or rejected depending on
the following criterion. By forming the quantity
\begin{equation}
a(\vectheta_\mathrm{trial}|\vectheta_i) = 
\min \left(1,\,\frac{p (\vectheta_{\mathrm{trial}})}{p (\vectheta_i)}
\frac{q(\vectheta_i| \vectheta_{\mathrm{trial}})}
{q(\vectheta_{\mathrm{trial}}| \vectheta_i)}\right),
\label{eq:acceptance}
\end{equation}
the trial state is accepted as a new state with a probability $a$ (rejected
with probability $1-a$). The transition 
probability is then 
\begin{equation}
\mathcal{T}(\vectheta_{i+1}|\vectheta_i) = 
a(\vectheta_\mathrm{trial}| \vectheta_i)q(\vectheta_{\mathrm{trial}}| \vectheta_i).
\label{eq:transition}
\end{equation}
If accepted, $\vectheta_{i+1} = \vectheta_{\mathrm{trial}}$, whereas if rejected,
the new state is equivalent to the current state, $\vectheta_{i+1} = \vectheta_i$.
This criterion ensures that once at its equilibrium, the chain samples the target 
distribution $p (\vectheta)$.
If the proposal density $q(\vectheta_{\mathrm{trial}}| \vectheta_i)$
is chosen to be symmetric, it cancels out in the expression of the acceptance probability,
which becomes: 
\begin{equation}
a = \min \left(1,\,\frac{p (\vectheta_{\mathrm{trial}})}
{p (\vectheta_i)}\right).
\label{eq:proposal_sym}
\end{equation}

We note that the process requires only evaluations of ratios of the target PDF.
This is a major virtue of this algorithm, in particular for Bayesian
applications, in which the normalisation factor in Eq.~(\ref{eq:bayes_theorem}), 
$P(\mathrm{data}) = \int P(\mathrm{data}|\vectheta) \cdot P(\vectheta) \mathrm{d} \vectheta $
 is often extremely difficult to compute. Hence, the ratio of the target PDF, i.e. the 
posterior of the parameter for our problem, can be calculated directly from the likelihood
 of the data and the priors.

        \subsection{Chain analysis}
        \label{sec:chain_analysis}
The chain analysis refers to the study of several properties
of the chains. The following quantities are inspected in order
to convert the chains in PDFs.

                \paragraph{Burn-in length}

The burn-in describes the practice of removing some
iterations at the beginning of the chain to eliminate the
{\em transient} time needed to reach the equilibrium or stationary
distribution, i.e., to {\em forget} the starting point. The burn-in
length $b$ is defined to be the number of first samples
$\{\vectheta_i\}_{i=1, \ldots, b}$ of the chain that must be discarded.
The stationary distribution is reached when the
chain enters the most probable parameter region corresponding to
the region where the target function is close to its maximal value.
To estimate $b$, the following criterion is used: we define
${p}_{1/2}$ to be the median of the target function distribution obtained from
 the entire chain of $N$ samples. The burn-in length
$b$ corresponds to the first sample $\vectheta_b$, for which 
${p} (\vectheta_b)\!>\!{p}_{1/2}$ (see App.~\ref{s:illustration} for an illustration).

                \paragraph{Correlation length}
By construction [see Eq.~(\ref{eq:transition})], each step of the chain depends
on the previous one, which ensures that the steps of the chain
are correlated. We can obtain independent samples by thinning the chain,
i.e. by selecting only a fraction of the steps 
with a periodicity chosen to derive uncorrelated samples. This period is estimated 
by computing the autocorrelation functions for each parameter.
For a parameter $\theta^{(\alpha)}$ ($\alpha = 1, \ldots, m$), the autocorrelation function
is given by
\begin{equation}
c_j^{(\alpha)} = \frac{E \left[ \theta_i^{(\alpha)} \theta_{j +
i}^{(\alpha)} \right] -
\left( E \left[ \theta_i^{(\alpha)} \right] \right)^2}{E \left[ \left(
\theta_i^{(\alpha)} \right)^2\right]},
\end{equation}
which we calculate with the Fast Fourier Transformation (FFT).
The correlation length $l^{(\alpha)}$ for the
$\alpha$-th parameter is defined as the smallest $j$ for
which $c_j^{(\alpha)}<1/2$, i.e. the
values $\theta^{(\alpha)}_i$ and $\theta^{(\alpha)}_{i+j}$
of the chain that are considered to be uncorrelated.
The correlation length $l$ for the chain, for all parameters,
is defined to be
\begin{equation}
  l \equiv \max_{\alpha = 1, \ldots, m} l^{(\alpha)},
\label{eq:correl_length}
\end{equation}
which is used as the period of the thinning (see App.~\ref{s:illustration}
for an illustration).

                \paragraph{Independent samples and acceptance}
The independent samples of the chain are chosen to be $\{\vectheta_i\}_{i=b+l k}$,
where $k$ is an integer. The number of independent samples $N_{\mathrm{ind}}$
is defined to be the fraction of steps remaining after discarding the burn-in steps
and thinning the chain,
\begin{equation}
N_{\mathrm{ind}}=\frac{N_{\mathrm{tot}}-b}{l}.
\label{f_ind1}
\end{equation}
 The independent acceptance $f_{\mathrm{ind}}$ is the
ratio of the number of independent samples $N_{\mathrm{ind}}$ to the total step
number $N_{\mathrm{tot}}$,
\begin{equation}
f_{\mathrm{ind}} = \frac{N_{\mathrm{ind}}}{N_{\mathrm{tot}}}.
\label{f_ind2}
\end{equation}

        \subsection{Choice of the target and trial functions}

                \subsubsection{Target function}

As already said, we wish to sample the target function 
$p(\vectheta)= P (\vectheta|\mathrm{data})$.
Using Eq.~(\ref{eq:bayes_theorem}) and the fact that the algorithm
is insensitive to the normalisation factor, this amounts to sampling the product
$P (\mathrm{data}|\vectheta) \cdot P (\vectheta)$. Assuming a flat prior
$P (\vectheta)={\rm cst}$, the target distribution reduces to
\begin{equation}
p (\vectheta)= P (\mathrm{data}|\vectheta)\equiv {\cal L}(\vectheta),
\label{eq:target}
\end{equation}
and here, the likelihood function is taken to be
\begin{equation}
{\cal L} (\vectheta)= \exp{\left( -\frac{\chi^2(\vectheta)}{2}\right)}.
\label{eq:likelihood}
\end{equation}
The $\chi^2(\vectheta)$ function for $n_{\rm data}$ data is
\begin{equation}
\chi^2 (\vectheta) = \sum_{k = 1}^{n_{\rm data}} \frac{(y^{\rm exp}_k -
y^{\rm theo}_k (\vectheta))^2}{\sigma_k^2},
\label{eq:chi2}
\end{equation}
where $y^{\rm exp}_k$ is the measured value, $y^{\rm theo}_k$
is the hypothesised value for both a certain model and the parameters
$\vectheta$, and $\sigma_k$ is the known variance of the measurement.
For example, $y^{\rm exp}_k$ and $y^{\rm theo}_k$ represent
the measured and calculated B/C ratios.

The link between the target function, i.e., the posterior PDF of the parameters,
and the experimental data is established with the help of
Eqs~(\ref{eq:target}) to (\ref{eq:chi2}). This link guarantees the proper
sampling of the parameter space using Markov chains, which spend
more time in more relevant regions of parameter space, as described above.

                \subsubsection{Trial function}
\label{s:trial_functions}

Despite the effectiveness of the Metropolis-Hastings
algorithm, to optimise the efficiency of the MCMC and minimise
the number of chains to be processed, trial functions should be
as close as possible to the true distributions.
We use a sequence of three trial functions to explore the
parameter space. The first step is a coarse
determination of the parameter PDF. This allows us to calculate
the covariance matrix leading to a better coverage of parameter space,
provided that the target PDF is sufficiently close to being an N-dimensional
Gaussian. The last step takes advantage of a binary-space partitioning (BSP)
algorithm. 

                        \paragraph{Gaussian step}
For the first iteration, the proposal density $q\left(\vectheta_{\rm trial},
\vectheta_i\right)$, required to obtain the trial value $\vectheta_{\rm trial}$
from $\vectheta_i$ is written as
\begin{equation}
q \left(\vectheta_{\rm trial}, \vectheta_i \right) \propto \prod_{\alpha=1,\ldots, m}
\exp{\left( - \frac{1}{2} \frac{\left(\theta^{(\alpha)}_{\rm trial} - \theta^{(\alpha)}_i
\right)^2}{\sigma_\alpha^2}\right)}.
\label{eq:gaussian}
\end{equation}
These represent $m$ independent Gaussian distributions centred on
$\vectheta_i$. The distribution is symmetric, so that the acceptance
probability $a$ follows Eq.~(\ref{eq:proposal_sym}).
The variance $\sigma_\alpha^2$ for each parameter $\alpha$
is to be specified. 
Each parameter $\theta_{\rm trial}^{(\alpha)}$ is hence calculated to be 
$$
\theta_{\rm trial}^{(\alpha)} = \theta_i^{(\alpha)} + \sigma_\alpha\cdot x,
$$
where $x$ is a random number obeying a Gaussian distribution centred on zero
with unit variance.

It is important to choose an optimal width $\sigma_\alpha$
to sample properly the posterior (target) distribution.
If the width is too large, as soon as the chain reaches a region of
high probability, most of the trial parameters fall into
a region of low probability and are rejected, leading to
a low acceptance and a long correlation length.
Conversely, for too small a width, the chain will take a longer time
to reach the interesting regions. Eventually, even if the chain
reaches these regions of high acceptance, only a partial
coverage of the PDF support will be sampled (also leading to
a long correlation length).

In practice, we first define $\sigma_\alpha$ ($\alpha=1, \ldots, m$) 
equal to the expected range of the parameter. In a subsequent iteration,
$\sigma_\alpha$ is set to be $2 \sqrt{2 \ln 2}\approx 2.3$ times
$\sigma_\alpha^{\rm calc}$, i.e. the FWHM of the PDF obtained with the
first iteration. The result is actually insensitive to the numerical factor used.

                        \paragraph{Covariance matrix} 

The proposal density is taken to be an N-dimensional Gaussian of covariance
matrix $V$
\begin{equation}
q(\vectheta_{\rm trial}, \vectheta_{i}) \!\propto \!\exp{\!\left(\!\!- \frac{1}{2}
(\vectheta_{\rm trial} - \vectheta_{i}) ^T V^{-1} (\vectheta_{\rm trial} - \vectheta_{i})
\!\right)}.
\label{eq:CM}
\end{equation}
The covariance matrix $V$ is symmetric and diagonalisable ($D$ is
a diagonal matrix of eigenvalues and $P$ represents the change in the coordinate matrix),
$$
V=P^T D P,
$$
and where again Eq.~(\ref{eq:proposal_sym}) holds.
The parameters $\vectheta_{\rm trial}$ are hence found to be
$$
\vectheta_{\rm trial} = \vectheta_i + P^T D \, \vec{x},
$$
where $\vec{x}$ is a vector of $m$ random numbers following
a Gaussian distribution centred on zero and with unit variance.

The covariance matrix $V$ is estimated, e.g., from a
previous iteration using the Gaussian step. The advantage of this 
trial function with respect to the previous one is
that it takes account of the possible correlations
between the $m$ parameters of the model.

       \paragraph{Binary Space Partitioning (BSP)} 
A third method was developed to define a proposal density for which the
results of the Gaussian step or the covariance matrix iterations are used to
subdivide the parameter space into boxes, in each of which a
given probability is affected. 

The partitioning of the parameter space can 
be organised using a binary-tree data structure known as a binary-space
partitioning tree \citep{BSP}.
The root node of the tree is the $m-$dimensional box corresponding 
to the entire parameter space. The binary-space partitioning is then performed 
by dividing each box recursively into two child boxes if the 
partitioning satisfies the following requirement: a box is divided 
only if the number of independent samples contained in this 
box is higher than a certain number (here we used a maximum of between 3\% and 
0.1\% of the total number of independent samples). When a box has to be 
divided, the division is made along the longer side of the box (the  box-side
lengths are defined  relative to  the root-box sides).
For each end node (i.e. node without any children), a probability, defined 
as the fraction of the number of independent samples in the box to their 
total number, is assigned. For empty boxes, a minimum 
probability is assigned and all the probabilities are renormalised so 
that the sum of all end-node probabilities equals 1. 

The proposal density  $q(\vectheta_{\rm{trial}})$ is then defined, in each
end-node box, as a uniform function equal to the assigned probability.
The sampling of this proposal density is simple and efficient: 
an end node is chosen with the assigned probability 
and the trial parameters are chosen uniformly in the corresponding box.
In comparison to the other two proposal densities, this
proposal density based on a BSP is asymmetric, because it is
only dependent on the proposal state $q(\vectheta_{\rm{trial}})$.
Hence, Eq.~(\ref{eq:acceptance}) must be used.

\section{Implementation in the propagation model}
\label{s:implementation}

The MCMC with the three above methods 
are implemented in the USINE package\footnote{A public version will
be released soon (Maurin, in preparation).}, which 
computes the propagation of Galactic CR nuclei and anti-nuclei for
several propagation models (LBM, 1D and 2D
diffusion models). The reader is referred to \citet{2001ApJ...555..585M}
for a detailed description for the nuclear parameters (fragmentation and
absorption cross-sections), energy losses (ionisation and Coulomb), and
solar modulation (force-field) used.

We briefly describe how the MCMC algorithm is implemented in the propagation part
(Sect.~\ref{s:implem}), using a LBM|the procedure would
be similar for any other model.
The LBM and its parameters are briefly discussed (Sect.~\ref{s:LBM}) as well
as the input spectrum parameters (Sect.~\ref{s:abundances}).
Additional information about the data are gathered in App.~\ref{s:config_data}.

	\subsection{Flow chart}
	\label{s:implem}

\begin{figure}[!t]
\centering
\includegraphics[width = 0.5\textwidth]{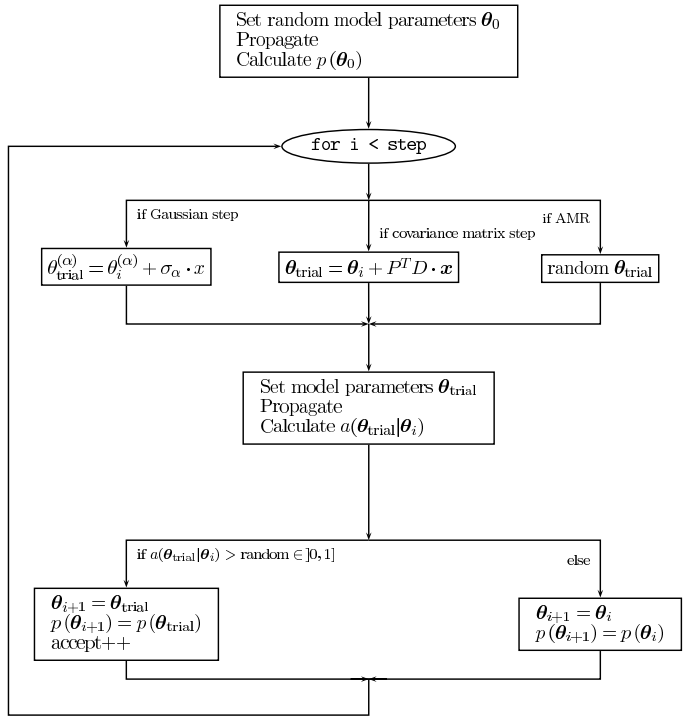}
\caption{Flow chart of the implemented MCMC algorithm:
$\vectheta_i$ is a vector [Eq.~(\ref{eq:theta})] of the
$\alpha=1, \ldots, m$ free parameters of the model, evaluated
at each step $i$, and $p (\vectheta_i)$ is the target function given
by Eq.~(\ref{eq:target}). See text for details.}
\label{fig:chart}
\end{figure}

A flow chart of the Metropolis-Hastings MCMC algorithm used
in the context of GCRs is given in Fig.~\ref{fig:chart}. 
To summarise, the initial values of the propagation parameters $\vectheta_{0}$ 
are chosen randomly in their expected range to crank up each Markov chain.
The interstellar (IS) CR fluxes are then calculated for this set
of parameters (see, e.g., Fig.~1 in
\citealt{2002A&A...394.1039M} for further details of the propagation steps).
The IS flux is modulated with the force-field approximation and the 
resulting top-of-atmosphere (TOA) spectrum is compared with the data,
which allows us to calculate the $\chi^2$ value [Eq.~(\ref{eq:chi2})], hence the likelihood
[Eq.~(\ref{eq:likelihood})]. This likelihood (in practice the log-likelihood)
is used to compute the acceptance probability [Eq.~(\ref{eq:acceptance})]
of the trial vector of parameters $\vectheta_{\rm trial}$ (as generated by
one of the three trial functions described in Sect.~\ref{s:trial_functions}). 
Whether the trial vector is accepted or rejected implies whether
$\vectheta_{1}=\vectheta_{\rm trial}$ or $\vectheta_{1}=\vectheta_{0}$.
This procedure is repeated for the $N$ steps of the chain. Obviously, 
when $\vectheta_{i+1}=\vectheta_{i}$, the propagation step does not
need to be repeated. Because of the nature of the MCMC algorithm, several chains can
be executed in parallel. Once completed, these chains are analysed (see
Sect.~\ref{sec:chain_analysis})|discarding the first step belonging
to the burning length and thinned according to the correlation length
$l$ [Eq.~(\ref{eq:correl_length})]|and combined to recover the
desired posterior PDF $P (\vectheta|\mathrm{data})$.

In this procedure, the user must decide i) the data to be used,  ii)
the observable to retain in calculating the likelihood, and iii) the number of
free parameters $m$ (of the vector $\vectheta$) for which we seek the posterior
PDF. 

	\subsection{Leaky-Box Model (LBM)\label{s:LBM}}

The LBM assumes that all CR species are confined 
within the Galaxy with an escape rate that equals $N/\tau_{\rm esc}$, where the escape 
time $\tau_{\rm esc}$ is rigidity-dependent, and is written as $\tau_{\rm esc}(R)$. 
This escape time has two origins.
First, CRs can leak out the confinement volume and leave the Galaxy.
Second, they can be destructed by spallation on interstellar matter nuclei.
This latter effect is parameterised by the grammage $x$ (usually expressed
in g~cm$^{-2}$), defined as the column density of 
interstellar matter encountered by a path followed by a CR.
The CRs that reach Earth have followed different paths,
and can therefore be described by a grammage distribution $N(x) \equiv dN/dx$.
The LBM assumes that 
\begin{equation}
N(x) \propto \exp^{-\lambda_{\rm esc}(R) x}\;,
\end{equation}
where the mean grammage $\lambda_{\rm esc}(R)= \langle x \rangle$ 
is related to the mass $m$, velocity $v$ and escape time
$\tau_{\rm esc}(R)$ by means of 
$\lambda_{\rm esc}(R) = \bar{m}nv \tau_{\rm esc}(R)$.

The function $\lambda_{\rm esc}(R)$ determines the amount of spallations
experienced by a primary species, and thus determines the secondary-to-primary 
ratios, for instance B/C. From an experimentalist point of view, $\lambda_{\rm
esc}(R)$ is a quantity that can be inferred from measurements of
nuclei abundance ratios. The grammage $\lambda_{\rm esc}(R)$ is known
to provide an effective description of diffusion
models~\citep{1990acr..book.....B}: it can be related to the
efficiency of confinement (which is determined by the diffusion coefficient and
to both the size and geometry of the diffusion volume), spallative destruction (which
tends to shorten the average lifetime of  a CR and thus lower
$\lambda_{\rm esc}$), and a mixture of other processes (such as convection,
energy gain, and losses).

In this paper, we compute the fluxes in the framework of the
LBM with minimal reacceleration by the interstellar
turbulence, as described in \citet{1988SvAL...14..132O} and \citet{1994ApJ...431..705S}.
The grammage $\lambda_{\rm esc}(R)$ is parameterised as
\begin{equation}\label{eq:lambdaesc}
 \lambda_{\rm esc}(R) = \begin{cases} \lambda_0 \beta R_0^{-(\delta-\delta_0)} R^{-\delta_0}&
\text{when $R < R_0$,}\\
\lambda_0 \beta R^{-\delta} &
\text{otherwise;}
\end{cases}
\end{equation}
where we allow for a break, i.e. a different slope below and above
a critical rigidity $R_0$. The standard form used in the literature
is recovered by setting $\delta_0=0$.
For the entire set of $n$ nuclei, a series of $n$ equations 
(see \citealt{2001ApJ...555..585M} for more details) for
the differential densities $N^{j=1,\ldots, n}$ are solved
at a given kinetic energy per nucleon $E_{k/n}$ ($E$ is the
total energy), i.e.
\begin{equation}
A^j N^j(E_{k/n}) + \frac{d}{dE}\left( B^j N^j - C^j \frac{dN^j}{dE} \right) = S^j(E_{k/n})\;.
\label{eq:transport_CR}
\end{equation}
In this equation, the r.h.s. term is the source term that takes into account the primary
contribution (see Sect.~\ref{s:abundances}), the spallative
secondary contribution from all nuclei $k$ heavier than $j$,
and the $\beta$-decay of radioactive nuclei into $j$. The
first energy-dependent factor $A^j$ is given by
\[
A^j = \frac{1}{\tau_{\rm esc}} 
      + \sum_{ISM=H,He} n_{\rm ISM} v^j \sigma^{j\rm +ISM}_{\rm inel}
		       + \frac{1}{\tau^j_\beta}.
\]
The two other terms correspond to energy losses and first-order
reacceleration for $B^j$ and to second-order reacceleration for $C^j$.
Following \citet{1988SvAL...14..132O} and \citet{1994ApJ...431..705S},
\[
B= \big\langle\frac{dE}{dt}\big\rangle_{\rm ion,\,coul.} +(1+\beta^2)\beta^2E K_{pp} \quad \text{and} \quad
C= \beta^4 E^2 K_{pp}\;,
\]
where
\begin{equation}
K_{pp}=\frac{4}{3} {\cal V}_a^2
 \frac{\tau_{\rm esc}}{\delta(4-\delta^2)(4-\delta)}\,.
\label{eq:Kpp}
\end{equation}
The strength of the reacceleration is mediated by the pseudo
Alfv\'enic speed ${\cal V}_a$ of the scatterers in units of km~s$^{-1}$~kpc$^{-1}$.
This is related to a true speed given in a diffusion model with a thin
disk $h$ and a diffusive halo $L$ by means of ${\cal V}_a=V_a\times (hL)^{-1/2}$
\citep{1994ApJ...431..705S}.
Assuming typical values of $h=0.1$~kpc and $L=10$~kpc, the value of
${\cal V}_a$ can be directly transposed and compared to a true speed $V_a$,
as obtained in diffusion models.

To summarise, our LBM with reacceleration may involve up to five
free parameters, i.e. the normalisation $\lambda_0$, the slopes
$\delta_0$ and $\delta$ below or above the cut-off rigidity $R_0$,
and a pseudo-Alfv\'en velocity ${\cal V}_a$ related to
the reacceleration strength.

	\subsection{Source spectra}
	\label{s:abundances}

We assume that the primary source spectrum $Q_j(E)$ for
each nuclear species $j$ is given by ($\beta=v/c$)
\begin{equation}
Q_j(E) \equiv dQ_j/dE = q_j \beta^{\eta_j} R^{- \alpha_j},
\label{eq:source_spec}
\end{equation}
where $q_j$ is the source abundance, $\alpha_j$ is the slope of the species $j$,
and the term $\beta^{\eta_j}$ manifests our ignorance about the low-energy spectral shape.
We further assume that $\alpha_j\equiv \alpha$ for all $j$,
and unless stated otherwise, $\eta_j\equiv \eta=-1$ in order to recover
$dQ/dp\propto p^{-\alpha}$, as obtained from acceleration models (e.g., \citealt{1994ApJS...90..561J}).
The constraints existing on $\eta$ are explored in Sect.~\ref{s:ModelIII+1}.

The pattern of the source abundances observed in the cosmic radiation
differs from that of the solar system. This is due to a segregation
mechanism during the acceleration stage. Two hypotheses are disputed
in the literature: one is based on the CR composition 
controlled by volatility and mass-to-charge ratio
\citep{1997ApJ...487..182M,1997ApJ...487..197E}, and the other one
is based on the first ionisation potential (FIP) of nuclei (e.g.,
\citealt{1973ICRC....1..584C}). In this work, for each configuration,
the source abundances are initialised
to the product of the solar system abundances \citep{2003ApJ...591.1220L},
and the value of the FIP taken from~\citet{1989ApJ...346..997B}. The final
fluxes are obtained by an iterative calculation of the propagated fluxes,
rescaling the element abundances|keeping fixed the relative isotopic abundances|to match experimental
data at each step until convergence is reached (see Fig.~1 in \citealt{2002A&A...394.1039M}
for further details). The result is thus insensitive to the input
values (more details about the procedure are given in App.~\ref{s:config_dataBC}).

The measurement of all propagated isotopic fluxes should
characterise all source spectra parameters completely, i.e.
the $q_j$ and $\alpha_j$ parameters should be free.
However, only element fluxes are available, which motivates the above
{\em rescaling} approach. In Sect.~\ref{s:ModelIII+1},
a few calculations are undertaken to determine self-consistently,
along with the propagation parameters, i) $\alpha$ and $\eta$, and ii)
the source abundances for the primary species C, O, and the mixed N elements
(the main contributors to the boron flux).

\section{Results\label{s:results}}

We first examine the relative merits of four different parameterisations
of the LBM, and determine the statistical significance of adding more parameters. These models
correspond to $\{\vectheta^{\alpha}\}_{\alpha=1, \ldots, m\leq5}$ with
\begin{itemize}
  \item Model~I~$=\{\lambda_0,\, R_0, \delta\}$, i.e. no reacceleration (${\cal V}_a = 0$) and no break in the spectral index ($\delta_0=0$).
  \item Model~II~$=\{\lambda_0,\, \delta, \, {\cal V}_a\}$, i.e. no critical rigidity ($R_0 = 0$) and no break in the spectral index ($\delta_0=0$).
  \item Model~III~$=\{\lambda_0,\, R_0, \, \delta, \, {\cal V}_a\}$, i.e. no break in the spectral index ($\delta_0=0$).
  \item Model~IV~$=\{\lambda_0,\, R_0, \, \delta_0, \, \delta, \, {\cal V}_a\}$.
\end{itemize}
Various subsets of B/C data are used to investigate
whether old data are useful or just add confusion to the PDF determination. 
We note in Sect.~\ref{s:pbar} that no useful constraint can be drawn from
\pbar\ data alone.

We also consider additional free parameters (Sect.~\ref{s:ModelIII+1}) related to the source spectra,
for a self-consistent determination of the propagation and source properties.
Since we show that a break in the slope (Model~IV) is not required by current data,
we focus on Model~III (for the description of the propagation parameters),
defining:
\begin{itemize}
	\item Model~III+1~$=\{\lambda_0,\, R_0, \, \delta, \, {\cal V}_a\}+ \{\alpha \}$, where the source slope $\alpha$ is a free parameter.
	\item Model~III+2~$=\{\lambda_0,\, R_0, \, \delta, \, {\cal V}_a\}+ \{\alpha, \eta \}$, where both the source slope $\alpha$ and the exponent $\eta$ [of $\beta$, see Eq.~(\ref{eq:source_spec})]
  are free parameters.
	\item Model~III+4~$=\{\lambda_0,\, R_0, \, \delta, \, {\cal V}_a\}+ \{\alpha,\, q_\mathrm{C}, \, q_\mathrm{N}, \, q_\mathrm{O} \}$, where the abundances $q_{i}$
  of the most significantly contributing elements are also free parameters.
	\item Model~III+5~$=\{\lambda_0,\, R_0, \, \delta, \, {\cal V}_a\}+ \{\alpha,\, \eta,\, q_\mathrm{C}, \, q_\mathrm{N}, \, q_\mathrm{O} \}$.
\end{itemize}
This allows us to investigate the correlations between parameters further and into
potential biases in the propagation parameter determination.

More details about the practical use of the trial functions
can be found in App.~\ref{s:illustration}. In particular, the sequential use
of the three sampling methods (Gaussian step, covariance matrix step, and then
binary-space partitioning) is found to be the most efficient: all
results presented hereafter are based on this sequence.

	\subsection{Fitting the B/C ratio\label{s:B/C}}

			\subsubsection{HEAO-3 data alone\label{s:HEAO}}
			
We first constrain the model parameters with HEAO-3 data only \citep{1990A&A...233...96E}.
These data are the most precise data available at the present day for the stable
nuclei ratio B/C of energy between 0.62 to \unit[35]{GeV/n}.

\begin{figure}[t]
\centering
\includegraphics[width = 0.47\textwidth]{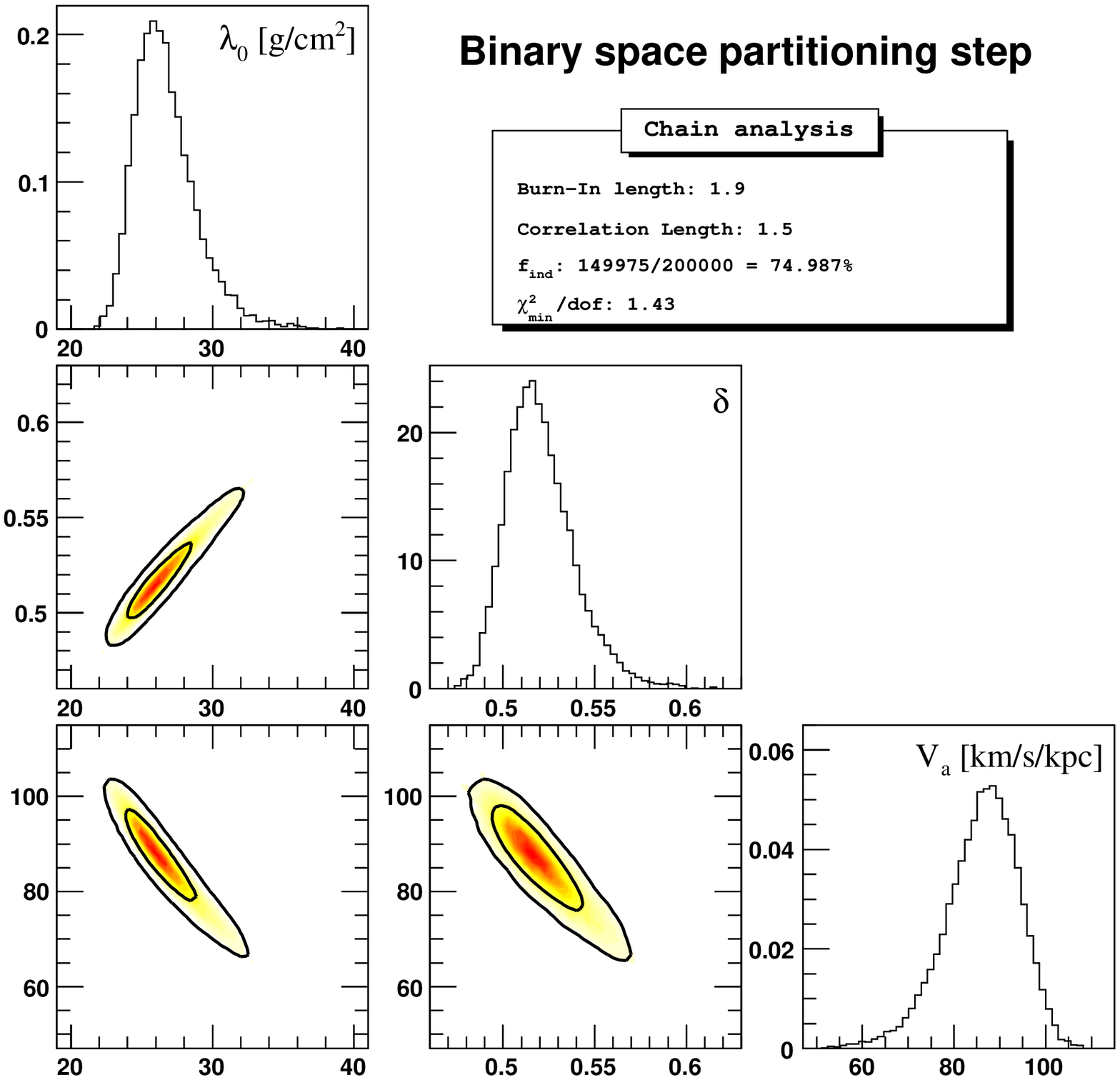}\vspace{3mm}
\includegraphics[width = 0.47\textwidth]{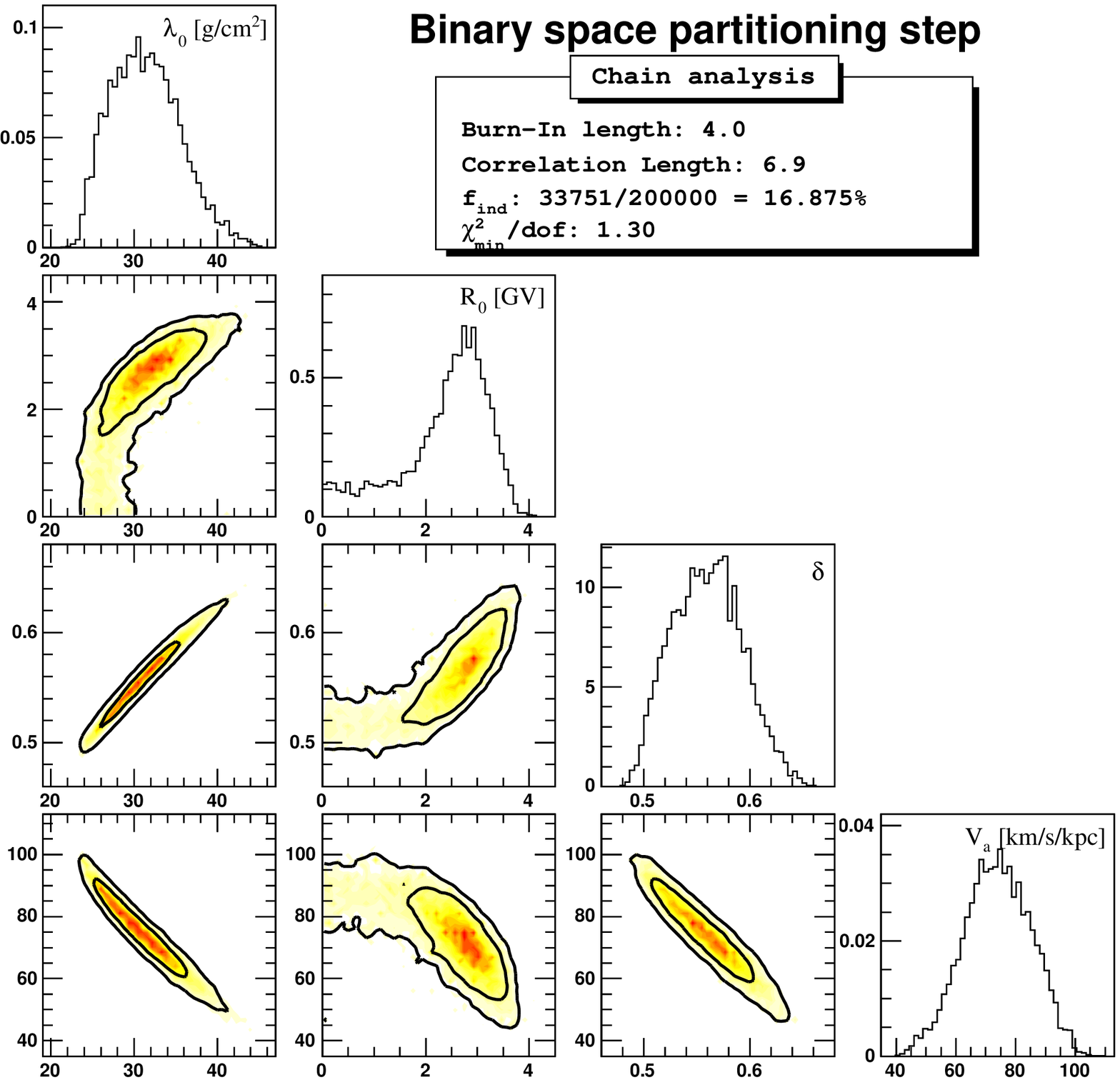}
\caption{Posterior distributions for Model II (top) and Model III (bottom) using HEAO-3 data only. For
more details, refer to caption of Fig.~\ref{fig:B2CwoVa}.}
\label{fig:Model_IIandIII}
\end{figure}

The results for the models I, II, and III are presented in Figs.~\ref{fig:B2CwoVa},
\ref{fig:Model_IIandIII} top, and 2 bottom.
The inner and outer contours are taken to be regions containing 68\% and 95\%
of the PDF respectively (see App.~\ref{s:CLparam}).
The first observation one can make for the LBM without reacceleration (Model I, 
Fig.~\ref{fig:B2CwoVa}),
is that the marginal distributions of the three LBM parameters are mostly Gaussian. 
The tail for small values of $R_0$ is due to this parameter being 
constrained by low-energy data ($ < \unit[1]{GeV/n}$): there are no HEAO-3 data 
at low energy, so all $R_0$ values below \unit[3]{GV} are equiprobable (this remains true
for Model III).

As seen in Fig.~\ref{fig:Model_IIandIII}, a more complicated shape for the different
parameters is found for Model II (top panel), and even more so for Model III (bottom panel).
This induces a longer correlation length (1.5 and 6.9 steps instead of 1 step) and
hence reduces the efficiency of the MCMC (75\% for model II and 17\% for model III).
Physically, the correlation between the parameters, as seen most clearly in Fig.~\ref{fig:Model_IIandIII} (bottom),
is understood as follows. First, $\lambda_0$, $R_0$, and $\delta$ are positively correlated.
This originates in the low-energy relation $\lambda_{\rm esc}\propto \lambda_0 R_0^{-\delta}$,
which should remain approximately constant to reproduce the bulk of the data at GeV/n energy.
Hence, if $R_0$ or $\delta$ is increased, $\lambda_0$
also increases to balance the product. On the other hand, ${\cal V}_a$ is negatively correlated
with $\delta$ (and hence with all the parameters): this is the standard result that
to reach smaller $\delta$ (for instance to reach a Kolmogorov spectrum), more
reacceleration is required. This can also be seen from Eq.~(\ref{eq:Kpp}),
where at constant $\tau_{\rm esc}$, $K_{pp}\propto {\cal V}_a^2/f(\delta)$, where
$f$ is a decreasing function of $\delta$: hence, if $\delta$ decreases,
$f(\delta)$ increases, and ${\cal V}_a$ then has to increase to retain the balance.

The values for the maximum of the PDF for the propagation parameters
along with their 68\% confidence intervals (see App.~\ref{s:CL}) are listed
in Table~\ref{tab:resultsHEAO}.
\begin{table}[t]
\centering
\begin{tabular}{cccccc} \hline\hline
Model & $\lambda_{0}$   &      $R_0$          &         $\delta$        &     ${\cal V}_a$       &  $\chi_{\mathrm{min}}^2/$dof \\ 
      &    g cm$^{-2}$  &       GV            &                         & $\!\!\!\!\!\!\!\!\!$km s$^{-1}$kpc$^{-1}\!\!\!\!\!\!\!\!\!$  & \\\hline
& \multicolumn{5}{c}{} \\ 
I     & $54^{+2}_{-2}$  & $4.2^{+0.3}_{-0.9}$ &  $0.70^{+0.01}_{-0.01}$ &        -         & 3.35 \vspace{0.1cm}\\
II    & $26^{+2}_{-2}$  &          -          &  $0.52^{+0.02}_{-0.02}$ & $88^{+6}_{-11}$  & 1.43\vspace{0.1cm}\\
III   & $30^{+5}_{-4}$  & $2.8^{+0.6}_{-0.8}$ &  $0.58^{+0.01}_{-0.06}$ & $75^{+10}_{-13}$ & 1.30 \vspace{0.1cm}\\\hline
\end{tabular}
\caption{Most probable values of the propagation parameters 
(after marginalising over the other parameters) for models I, II, and III
using HEAO-3 alone (14 data points) and the B/C constraint.  The uncertainty in the
parameters correspond to 68\% CL of the marginalised PDF (see App.~\ref{s:CL}). The last column shows the minimum $\chi^2/$dof
obtained for each model (the associated best-fit parameters are gathered in Table~\ref{tab:resultsBestFit}).}
\label{tab:resultsHEAO}
\end{table}
The values obtained for our Model~I are in fair agreement with those derived
by~\citet{1998ApJ...508..940W}, who found $\{\lambda_0,\, R_0,\, \delta \}=\{
38.27,\, 3.6,\, 0.7\}$. The difference for $\lambda_0$ could be related to the
fact that \citet{1998ApJ...508..940W} rely on a mere eye inspection to extract
the best-fit solution or/and use a different set of data. For example, comparing
Model~I with a combination of HEAO-3 and low-energy data (ACE+Voyager$\,$1$\,$\&$\,$2+IMP7-8, see
Sect.~\ref{sec:lowenergydata}) leads to $\{\lambda_0,\, R_0,\, \delta \}=\{52,\, 5.3,\,
0.69\}$, slightly changing the values of the Model~I preferred parameters
(compared to the first line of Table~\ref{tab:resultsHEAO}).

The reacceleration mechanism was invoked in the literature to decrease the spectral index $\delta$
toward its preferred value of $1/3$	 given by a Kolmogorov spectrum of turbulence. In Table~\ref{tab:resultsHEAO},
the estimated propagation parameter values for the models II and III are indeed slightly smaller than
for Model~I, but the Kolmogorov spectral index is excluded for all of these three cases (using HEAO-3 data only).
This result agrees with the findings of \citet{2001ApJ...555..585M}, in which a more realistic
two-dimensional diffusion model with reacceleration and convection was used. 
We note that the values for ${\cal V}_a\sim80$~km~s$^{-1}$~kpc$^{-1}$, should lead to
a true speed $V_a={\cal V}_a\times \sqrt{hL}\sim80$~km~s$^{-1}$ in a diffusion
model for which the thin disk
half-height is $h=0.1$~kpc and the halo size is $L=10$~kpc: this is
consistent with values found in \citet{2002A&A...394.1039M}.

\begin{figure}[t]
\centering
\includegraphics[width = .5\textwidth]{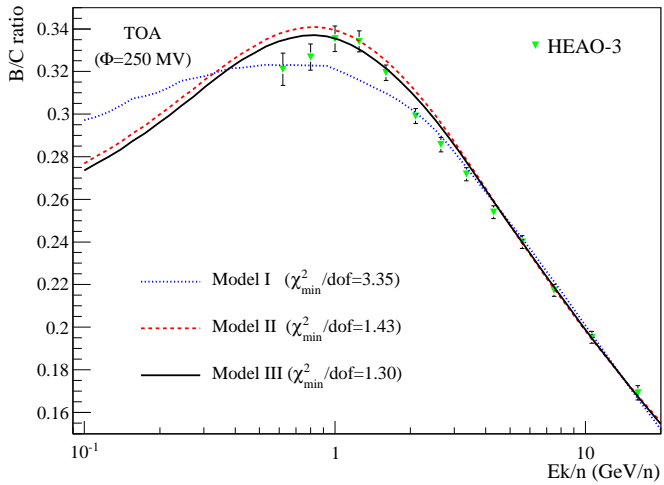}
\caption{Best-fit ratio for Model I (blue dotted), II (red dashed), and
Model III (black solid) using the HEAO-3 data only (green symbols). The curves are
modulated with $\Phi = \unit[250]{GV}$. The corresponding best-fit parameters are gathered
in Table~\ref{tab:resultsBestFit}.}
\label{fig:Models_I_II_III}
\end{figure}

The final column in Table~\ref{tab:resultsHEAO} indicates, for each model, the best $\chi^2$
value per degree of freedom, $\chi_{\mathrm{min}}^2/$dof. This allows us to compare the relative merit of the models.
LB models with reacceleration reproduce the HEAO-3 data more accurately (with $\chi^2$/dof of 1.43 and 1.30 for the Models
II and III respectively compared to $\chi^2$/dof = 4.35 for Model~I). The best-fit model B/C fluxes are shown with the
B/C HEAO-3 data modulated at $\Phi=250$~MV in Fig.~\ref{fig:Models_I_II_III}.
Physically, the origin of a cutoff $R_0$ in $\lambda_{\rm esc}$
at low energy can be related to convection in diffusion models \citep{1979ApJ...229..747J}. Hence, it
is a distinct process as reacceleration.
The fact that Model III performs more successfully than Model II implies that both processes
are significant, as found in~\citet{2001ApJ...555..585M}.

In the following, we no longer consider Model I and II, and inspect instead, the parameter
dependence of Model III on the dataset selected.

			\subsubsection{Additional constraints from low-energy data}
			\label{sec:lowenergydata}

The actual data sets for the B/C ratio (see, e.g., Fig.~\ref{fig:Models_III_A_C})
show a separation into two energy domains: the low-energy 
range extends from \unit[$\sim10^{-2}$]{GeV/n} to \unit[$\sim 1$]{GeV/n} 
and the high-energy range goes from \unit[$\sim 1$]{GeV/n} to 
\unit[$\sim 10^{2}$]{GeV/n}. The spectral index $\delta$ is
constrained by high-energy data, e.g., the HEAO-3 data,
and adding low-energy data allows us to more reliably constrain $R_0$.
We note that by fitting only the low-energy data, only the grammage
crossed in very narrow energy domain would be constrained. 

In a first step, we add only the ACE (CRIS) data \citep{2006AdSpR..38.1558D}, which covers the energy range
from \unit[$\sim  8 \cdot 10^{-2}$]{GeV/n} to \unit[$\sim 2 \cdot 10^{-1}$]{GeV/n}, and which is later referred to as dataset B
(the dataset A being HEAO-3 data alone). The resulting posterior
distributions are similar for the datasets B and A (B is not shown, but A is given in Fig.~\ref{fig:Model_IIandIII}, bottom).
Results for datasets A and B are completely consistent (first and second line of Table~\ref{tab:resultsHEAO-lowE}),
but for the latter, propagation parameters are more tightly constrained and the fit is improved
($\chi^{2}_{\rm min}$/dof=1.09). The ACE (CRIS) data are compatible with $R_{0} = 0$, but
the preferred critical rigidity is \unit[2.47]{GV}.

\begin{figure}[t]
\centering
\includegraphics[width = .5\textwidth]{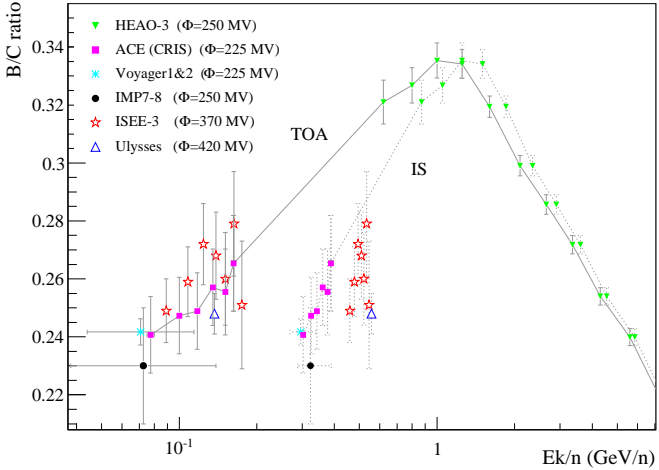}
\caption{HEAO-3 and ACE (CRIS) modulated (TOA, solid line) and demodulated (IS, dashed line) data points
have been connected to guide the eye. Filled symbols (modulated
and demodulated) correspond to HEAO-3, ACE (CRIS), IMP7-8 and Voyager$\,$1$\,$\&$\,$2. On the TOA curve,
the empty red stars and the blue upper triangle correspond to ISEE-3 and Ulysses.}
\label{fig:BC_mod_demod}
\end{figure}

\begin{figure}[t]
\centering
\includegraphics[width = .5\textwidth]{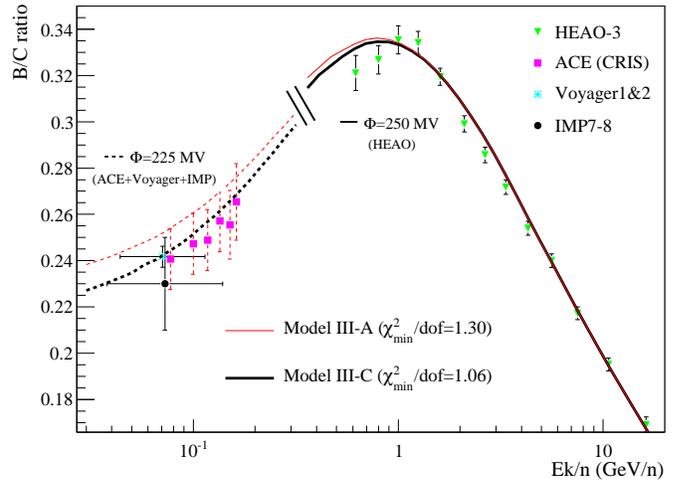}
\caption{Best-fit B/C flux (Model III) for datasets A (thin red curves) and C
(thick black curves). Above 300 MeV/n, B/C is modulated to $\Phi=250$~MV (solid lines)
appropriate for HEAO-3 data, whereas below, it is modulated to $\Phi=225$~MV (dashed lines).
Model III-B, not shown, overlaps with III-C. The corresponding propagation values are gathered in
Table~\ref{tab:resultsBestFit}.}
\label{fig:Models_III_A_C}
\end{figure}

All other low-energy data (ISEE-3, Ulysses, IMP7-8, Voyager$\,$1$\,$\&$\,$2, ACE) are then included (dataset D). 
The resulting values of the propagation parameters are left unchanged. However, a
major difference lies in the higher $\chi^{2}_{\rm min}$/dof of 4.15, which reflects an inconsistency
between the different low-energy data chosen for the MCMC. If the data point from the Ulysses experiment
is excluded,  $\chi^{2}_{\rm min}$/dof decreases to a value of 2.26, and
by excluding also the ISEE-3 data points (dataset C) it decreases further to 1.06
(see Table \ref{tab:resultsHEAO-lowE}).
\begin{table}[t]
\begin{center}
\begin{tabular}{ccccccc}\hline\hline
  Model        &   $\lambda_{0}$   &   $R_0$             &   $\delta$              &   ${\cal V}_a$          &   $\chi_{\mathrm{min}}^2/$dof \\
$\!\!\!$Dataset$\!\!\!$ & $\!$g cm$^{-2}\!\!\!$  &   GV           &                         & $\!\!\!\!\!\!$km s$^{-1}$kpc$^{-1}\!\!\!\!\!\!$  & \\\hline
& \multicolumn{5}{c}{}\\ 
   III-A       & $30^{+5}_{-4}$    & $2.8^{+0.6}_{-0.8}$ &  $0.58^{+0.01}_{-0.06}$ & $75^{+10}_{-13}$ & 1.30 \vspace{0.1cm}\\
   III-B       & $28^{+2}_{-3}$    & $2.6^{+0.4}_{-0.7}$ &  $0.53^{+0.02}_{-0.03}$ & $85^{+9}_{-8}$   & 1.09 \vspace{0.1cm}\\
   III-C       & $27^{+2}_{-2}$    & $2.6^{+0.4}_{-0.7}$ &  $0.53^{+0.02}_{-0.03}$ & $86^{+9}_{-5}$    & 1.06 \vspace{0.5cm}\\
   III-D       & $26^{+2}_{-2}$    & $3.0^{+0.4}_{-0.5}$ &  $0.52^{+0.02}_{-0.02}$ & $95^{+7}_{-6}$   & 4.15 \vspace{0.1cm}\\
   III-E       & $30^{+2}_{-2}$    & $3.7^{+0.2}_{-0.3}$ &  $0.57^{+0.01}_{-0.02}$ & $88^{+3}_{-6}$   & 6.08  \vspace{0.1cm}\\ \hline
\end{tabular}
\caption{Same as in Table~\ref{tab:resultsHEAO},
but testing different data sets with Model~III:
A = HEAO-3 (14 data points), B = HEAO-3 + ACE (20 data points), C = HEAO-3 + ACE + Voyager$\,$1$\,$\&$\,$2 + IMP7-8 
(22 data points), D = HEAO-3 + all low-energy data (30 data points), E = all B/C data (69 data points).
}
\label{tab:resultsHEAO-lowE}
\end{center}
\end{table}
Since the set of low-energy data have different modulation parameters, the 
difference in the results for the various data subsets becomes clearer after
the data have been demodulated. The force-field approximation
provides a simple analytical one-to-one correspondence between the modulated
top-of-the atmosphere (TOA) and the demodulated interstellar (IS) fluxes. For an isotope
$x$, the IS and TOA energies per nucleon are related by $E_k^{\rm IS} = E_k^{\rm TOA} +
\Phi$ ($\Phi=Z/A\times \phi$ is the modulation parameter), and the fluxes
by ($p_x$ is the momentum per nucleon of $x$)
\begin{equation}
\psi_x^{\rm IS} \left(E_k^{IS}\right) = \left( \frac{p_x^{\rm IS}}{p_x^{\rm TOA}}\right)^2 \psi_x^{\rm TOA} \left(E_k^{\rm TOA} + Z/A \times \psi\right).
\end{equation}
The B/C ratio results from a combination of various isotopes, and assuming
the same Z/A for all isotopes, we find that
\begin{equation}
\left( \frac{\rm B}{\rm C}\right)^{\rm IS} \left(E_k^{IS}\right) = \left( \frac{\rm B}{\rm C}\right)^{\rm TOA} \left(E_k^{\rm TOA} + Z/A \times \phi \right).
\end{equation}
The modulated and demodulated low-energy B/C data are shown in Fig.~\ref{fig:BC_mod_demod}
(see caption for details). The ISEE-3 and Ulysses data points, as just underlined, 
are clearly inconsistent with other data. To be consistent, $\Phi=200$~MV for
ISEE-3 and $\Phi=200$~MV for Ulysses would be required. Significant uncertainties
$\Delta\Phi\sim 25-50$~GV are quoted in general, so that it is difficult to
conclude whether there are systematics in the measurement or if the modulation quoted
in the papers is inappropriate. Some experiments have also accumulated the signal for
several years, periods during which the modulation changes. It is beyond the
scope of this paper to discuss this issue further.
Below, we discard both ISEE-3 and Ulysses data in selecting an homogeneous
low-energy data set, which includes the most recent ACE (CRIS) data.
\begin{table}[t]
\centering
\begin{tabular}{ccccccc} \hline\hline
Model    & $\lambda_{0}^{\rm best}$ & $\!R_0^{\rm best}\!$ & $\delta_{0}^{\rm best}$ & $\delta^{\rm best}$ &       ${\cal V}_a^{\rm best}$           & $\!\!\chi^2$/dof$\!\!$   \\
Data     & $\!\!$g cm$^{-2}\!\!$    &          GV          &                         &                     & $\!\!\!\!$km$\,$s$^{-1}$kpc$^{-1}\!\!\!\!$ & \\\hline
& \multicolumn{5}{c}{} \\ 
I-A      & 54.7 & 4.21 & - & 0.702 & -  & 3.35 \vspace{0.1cm}\\
II-A     & 25.8 & - & - & 0.514 & 88.8  & 1.43 \vspace{0.1cm}\\
III-A    & 31.7 & 2.73 & - & 0.564 & 73.0  & 1.30 \vspace{0.1cm}\\
III-C    & 26.9 & 2.45 & - & 0.527 & 88.5  & 1.06 \vspace{0.1cm}\\
IV-C     & 32.7 & 2.38 & -0.97 & 0.572 & 70.5 & 0.86 \vspace{0.3cm}\\
\hline
\end{tabular}
\caption{Best-fit values (corresponding to $\chi^2_{\rm min}$)
for B/C data (A=HEAO-3 data alone, C=HEAO-3+Voyager$\,$1$\,$\&$\,$2+ACE+IMP7-8).}
\label{tab:resultsBestFit}
\end{table}

The resulting best-fit models, when taking low-energy data into account, are
displayed in Fig.~\ref{fig:Models_III_A_C}. The B/C best-fit ratio is displayed for
Model III and the dataset A (red thin lines) and C (black thick lines). Model III-B
(not shown) yields similar results to Model III-C. Solid and dashed lines correspond
to the two modulations $\Phi = 250$~MV (HEAO-3 and IMP7-8) and $\Phi = 225$~MV 
respectively (ACE and Voyager$\,$1$\,$\&$\,$2). Although the fit from HEAO-3 alone provides a
good match at low energy, adding ACE (CRIS) and Voyager$\,$1$\,$\&$\,$2 constraints slightly
shifts all of the parameters to lower values.

In a final try, we take into account all available data (dataset E,
final line of Table~\ref{tab:resultsHEAO-lowE}).
Many data are clearly inconsistent with each other (see Fig.~\ref{fig:BC_envelopes}), but as
for the low-energy case, although the $\chi^{2}_{\rm min}$/dof
is worsened, the preferred values of the propagation parameters are not
changed drastically (compare with datasets B, C, and D in Table~\ref{tab:resultsHEAO-lowE}).
We await forthcoming data from CREAM, TRACER, and PAMELA to be able to confirm
and refine the results for HEAO-3 data.


\subsubsection{Model IV: break in the spectral index}

We have already mentioned that the rigidity cut-off
may be associated with the existence of a galactic wind
in diffusion models. By allowing a break to occur in the spectral
index of $\lambda_{\rm esc}$ [see Eq.~(\ref{eq:lambdaesc})],
we search for a deviations from a single power law
($\delta_0=\delta$) or from the cut-off case ($\delta_0=0$).

\begin{figure}[t]
\centering
\includegraphics[width = .5\textwidth]{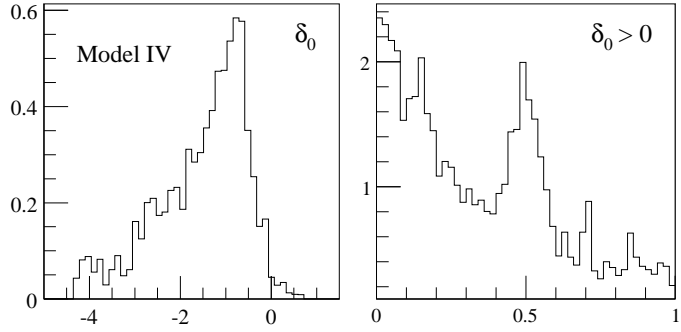}
\caption{Marginalised PDF for
the low-energy spectral index $\delta_0$ in Model~IV-C. The parameter $\delta_0$ is either
free to span both positive and negative values (left panel) or constrained to $\delta_0 > 0$ (right panel).}
\label{fig:PDF_delta0}
\end{figure}

Adding a new parameter $\delta_0$ (Model~IV) increases the correlation length of
the MCMC, since $R_{0}$ and $\delta_{0}$ are correlated [see Eq.~(\ref{eq:lambdaesc})].
The acceptance $f_{\mathrm{ind}}$ [Eq.~(\ref{f_ind1})] is hence extremely low. For Model IV-C
(i.e. using dataset C, see Table~\ref{tab:resultsHEAO-lowE}),
we find $f_{\mathrm{ind}}=2\%$. The PDF for $\delta_0$ is shown in the
left panel of Fig.~\ref{fig:PDF_delta0}. The most probable values and 68\%
confidence intervals obtained are $\{\lambda_0, R_0, \delta_0,
\delta, V_{a} \} = \{ 30^{+2}_{-2}, 2.2^{+0.4}_{-0.6}, -0.6^{+0.2}_{-1.3}, 0.55^{+0.04}_{-0.02}, 76^{+9}_{-11} \}$,
which are consistent with values found for other models,
as given in Tables~\ref{tab:resultsHEAO} and \ref{tab:resultsHEAO-lowE}:
adding a low-energy spectral break only allows us to better adjust low-energy
data (figure not shown). The best-fit parameters, for which $\chi^2_{\rm min}=0.86$,
are reported in Table~\ref{tab:resultsBestFit}. The small value of $\chi^2_{\rm min}$
(smaller than 1) may indicate an over-adjustment, which would disfavour the model.

It is also interesting to compel $\delta_0$ to be positive, to check whether $\delta_0=0$ (equivalent
to Model~III), $\delta_0=\delta$ (equivalent to Model~II), or any
value in-between that is preferred. We find the most probable values to be
$\{ \lambda_0, R_0, \delta_0, \delta, V_{a} \} = \{ 23^{+1}_{-1}, 1^{+2}_{-1},
0^{+0.6}, 0.49^{+0.01}_{-0.01}, 102^{+4}_{-5} \}$. The corresponding PDF for $\delta_0$
is shown in the right panel of Fig.~\ref{fig:PDF_delta0}. The maximum occurs for
$\delta_0=0$, which is also found to be the best-fit value; we checked that
the best-fit parameters matches those given in Table~\ref{tab:resultsBestFit} for Model~III-C.
A secondary peak appears at $\delta_0\approx 0.5$, such as $\delta_0\approx\delta$
corresponding to Model~II. The associated $\chi^2_{\rm min}$ for this configuration
is worse than that obtained with $\delta_0=0$, in agreement with the conclusion
that Model~III provides a closer description of the data than Model~II.


			\subsubsection{Summary and confidence levels for the B/C ratio}
			\label{sec:BCenvelopes}

In the previous paragraphs, we have studied several models and
B/C datasets. The two main conclusions that can be drawn
are i) the best-fit model is Model~III, which includes reacceleration
and a cut-off rigidity, and ii) the most likely values of the propagation parameters
are not too dependent on the data set used, although when data are inconsistent
with each other the statistical interpretation of the goodness of fit of a
model is altered (all best-fit parameters are gathered in 
Table~\ref{tab:resultsBestFit}). The values of the derived propagation parameters
are close to the values found in similar studies and the correlation between the LB
transport parameters are well understood.

\begin{figure}[t]
\centering
\includegraphics[width = .5\textwidth]{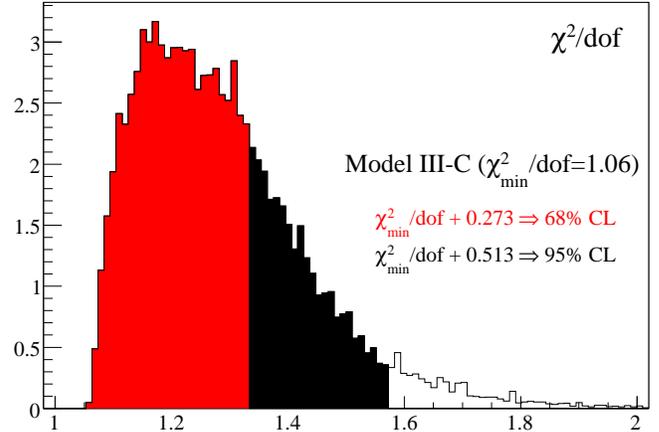}
\caption{$\chi^2$/dof normalised distribution for Model~III-C. The 68\% and 95\% CL
of the distribution are shown respectively as the red and black area.}
\label{fig:chi2_envelopes}
\end{figure}

Taking advantage of the knowledge of the $\chi^2$ distribution, we can extract
a list of configurations, i.e. a list of parameter sets, based on CLs of the $\chi^2$
PDF (as explained in App.~\ref{s:CLflux}). The $\chi^2$ distribution is shown for
our best model, i.e. Model~III, in Fig.~\ref{fig:chi2_envelopes}. The red and black areas correspond
to the 68\% and 95\% confidence intervals, which are used to generate two configuration lists,
from which 68\% and 95\% CLs on, e.g., fluxes, can be derived\footnote{For instance,
it can be used to predict the \pbar\ or $\bar{d}$ background flux to look for
a dark-matter annihilating contribution, e.g., as in \citet{2004PhRvD..69f3501D}.
Note however that the statistical procedure described here is far more robust than 
the crude approach used by \citet{2001ApJ...555..585M,2004PhRvD..69f3501D}.}.

\begin{figure}[t]
\centering
\includegraphics[width = .5\textwidth]{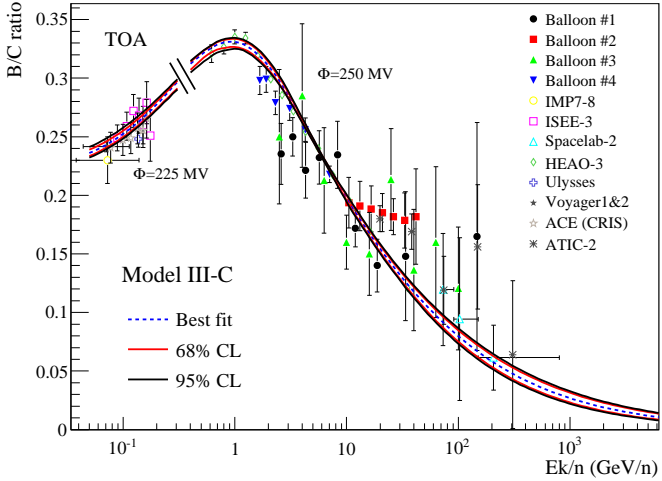}
\caption{Confidence regions of the B/C ratio for Model~III-C as calculated from all
propagation parameters satisfying Eq.~(\ref{eq:CI_PDF}). The blue-dashed line is the best-fit
solution, red-solid line is 68\% CL and black-solid line 95\% CL. Two modulation parameters
are used: $\Phi=225$~MV below 0.4 GeV/n (adapted for ACE+Voyager$\,$1$\,$\&$\,$2+IMP7-8 data)
and $\Phi=250$~MV above (adapted for HEAO-3 data).}
\label{fig:BC_envelopes}
\end{figure}

The B/C best-fit curve (dashed blue), the 68\% (red solid), and 95\% (black solid) CL
envelopes are shown in Fig.~\ref{fig:BC_envelopes}. 
For the specific case of the LBM, this demonstrates that current data are already
able to constrain strongly the B/C flux (as reflected by the
good value $\chi^2_{\rm min}=1.06$), even at high energy.
This provides encouraging support in the discriminating power of forthcoming data.
However, this conclusion must be confirmed by analysis of a more refined model
(e.g., diffusion model), for which the situation might not be so simple.

From the same lists, we can also derive the range allowed for the
source abundances of elements (we did not try to fit isotopic abundances here,
although this can be achieved, e.g., as in \citealt{2001SSRv...97..337S} and references therein).
The element abundances are gathered in Table~\ref{tab:abundances}, for elements
from C to Si (heavier elements were not used in this study). They can be compared
with those found in \citet{1990A&A...233...96E} (see also those derived from Ulysses
data, \citealt{1996A&A...316..555D}). For some elements, the agreement is striking (F, Mg),
and is otherwise fair. The difference for the main progenitors of boron, i.e. C, N, and O,
is a bit puzzling, and is probably related to a difference in the input-source spectral shape. 
This is discussed further in Sect.~\ref{s:ModelIII+1}, where we also determine
self-consistently the propagation parameters along with the C, N, and O abundances.
\begin{table}[t]
\begin{center}
\begin{tabular}{ccll}\hline\hline
  Z    & $\!\!$Element$\!\!$ &  $~~~10^{22}\times q_Z$ &  ~~~~~HEAO-3 \\
	     &                     &  ~(m$^3$~s~GeV/n)$^{-1}$ &  (Engelmann et al.) \vspace{0.5mm}\\\hline
	6    &          C          &  $148.5~~\pm~3.		      $   &  $164.9~~\pm~4.7		 $\\
	7    &          N          &  $~~~8.1~~\pm~0.6	      $   &  $~~~9.9~~\pm~3.4	   $\\
	8    &          O          &  $185.~~~\,\pm~3.        $   &  $204.~~~\,\pm~2.2  $\\
	9    &          F          &  $~~~3.67\,\pm~0.05      $   &  $~~~3.67\,\pm~0.05  $\\
	10   &          Ne         &  $~\,24.1~~\pm~0.4 	    $   &  $~\,22.5~~\pm~1.3 	 $\\
	11   &          Na         &  $~~~1.88\,\pm~0.08      $   &  $~~~1.25\,\pm~0.5   $\\
	12   &          Mg         &  $~\,40.3~~\pm~0.6       $   &  $~\,40.3~~\pm~1.0   $\\
	13   &          Al         &  $~~~3.69\,\pm~0.1       $   &  $~~~3.02\,\pm~0.6   $\\
	14   &          Si         &  $~\,{\bf 38.8}~~\pm~0.5 $   &  $~\,{\bf 38.8}~~\pm~0.5   $\\\hline
\end{tabular}
\caption{Element source abundances, $q_Z\equiv\sum_{i=\rm isot.} q_i$, for Model~III-C (isotopic fractions are
fixed to SS ones, \citealt{2003ApJ...591.1220L}). The central values correspond, for the best-fit model,
to abundances rescaled to match HEAO-3 data at 10.6 GeV/n. The uncertainty
in $q_i$ originates from the same rescaling, but arising from all combinations of
parameters satisfying the 68\% CL on the $\chi^2$ distribution. For HEAO-3, the numbers are taken
from Table~7 of \citet{1990A&A...233...96E}, and have been rescaled to
$q_Z({\rm Si})=38.8\times10^{-22}$~(m$^3$~s~GeV/n)$^{-1}$ to ease the comparison.}
\label{tab:abundances}
\end{center}
\end{table}
%

	\subsection{Constraints from \pbar \label{s:pbar}}

In the context of indirect dark-matter searches, the antimatter
fluxes (\pbar, $\bar{d}$ and $e^+$) are used to look for
exotic contributions on top of the standard, secondary ones. 

\begin{figure}[t!]
\centering
\includegraphics[width = .5\textwidth]{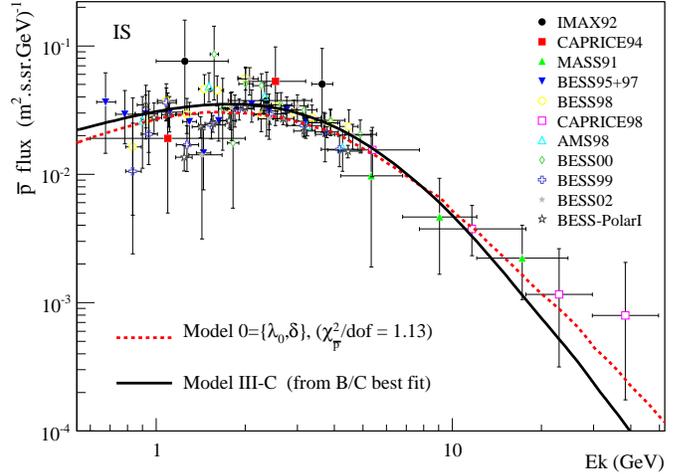}
\caption{Demodulated anti-proton data and IS flux for Model 0 (best fit on \pbar\ data,
red-dashed line) and for Model III-C (from the best-fit parameters on B/C data, black-solid line).}
\label{fig:flux_pbar}
\end{figure}

The standard procedure is to fit the propagation parameters
to B/C data, and apply these parameters in calculating the
secondary and primary (exotic) contributions.
The secondary flux calculated for our best-fit Model~III-C is
shown, along with the data (see App.~\ref{s:config_dataPbar} for
more details) in Fig.~\ref{fig:flux_pbar} (black-solid line).
For this model, we can calculate the $\chi^2$ value for the \pbar\
data, and we find $\chi^2/$dof=1.86. The fit is not perfect, and
as found in other studies (e.g., \citealt{2005PhRvD..71h3013D}),
the flux is somehow low at high energy (PAMELA data are awaited
to confirm this trend). However, we checked that these high-energy
data points are not responsible for the {\em large} $\chi^2$ value.
The latter could be attributed to either a small exotic contribution,
a different propagation history for species for which
$A/Z=1$ or $A/Z\approx2$, or inaccurate data.

It may therefore appear reasonable to fit directly the propagation parameters
to the \pbar\ flux, assuming that it is a purely secondary species.
Since the fluxes of its progenitors ($p$ and He) are well measured,
this should provide an independent check of the propagation history.
We first attempted to apply the MCMC method to the \pbar\ data with Model III,
then Model~II and finally Model~I. However, even the simplest model exhibits
strong degeneracies, and the MCMC chains could not converge. We had to revert
to a model with no reacceleration (${\cal V}_a = 0$), no critical rigidity
($R_0 = 0$), and no break in the spectral index ($\delta_0=0$), for which
$\lambda_{\rm esc}=\lambda_0 \beta (R/1GV)^{-\delta}$ (hereafter Model~0).
The $1\sigma$ values found for the two parameters $\{\lambda_0, \delta\}$
are $\lambda_0^{\bar{p},\,\rm Model~0}=\unit[10.2^{+0.5}_{-0.5}]{g \cdot
cm^{-2}}$ and $\delta^{\bar{p},\,\rm Model~0}=0.00^{+0.04}$.
Hence, only one parameter ($\lambda_0$) is required to reproduce
the data, as seen in Fig.~\ref{fig:flux_pbar} (red-dashed line,
$\chi^2_{\rm min}/$dof=1.128).
This is understood as follows: due to the combined effect of modulation and
the tertiary contribution (\pbar\ inelastically interacting on the ISM, but
surviving as a \pbar\ of lower energy), the true low-energy data points all
correspond to \pbar\ produced at a few GeV energies. Due to
the large scattering in the data, it is sufficient to produce the correct
amount of \pbar\ at this particular energy to account for all of the data.

Due to the importance of antimatter fluxes for indirect dark-matter
searches, this novel approach could be helpful in the future. However, this
would require a more robust statistics of the \pbar\ flux, especially
at higher energy, to lift the degeneracy in the parameters.

	\subsection{Adding free parameters related to the source spectra\label{s:ModelIII+1}}
In all previous studies (e.g., \citealt{2001ApJ...547..264J}), the source parameters were
investigated after the propagation parameters had been determined from the B/C ratio
(or other secondary to primary ratio). We propose a more general approach, where we
fit simultaneously all of the parameters. With the current data, this already provides strong
constraints on the CR source slope $\alpha$ and source abundances (CNO).
Higher-quality data are awaited to refine this analysis.
We also show how this approach can help to uncover inconsistencies in the measured fluxes.

For all models below, taking advantage of the results obtained in Sect.~\ref{s:B/C},
we retain Model~III-C. The roman number refers to the free transport parameters
of the model (III$=\{\lambda_0,\, R_0, \, \delta, \, {\cal V}_a\}$), and the capital refers to
the choice of the B/C dataset (C=HEAO-3+Voyager$\,$1$\,$\&$\,$2+IMP7-8, see Table~\ref{tab:resultsHEAO-lowE}).
This is supplemented by source spectra parameters and additional data for the element
fluxes.

\subsubsection{Source shape $\alpha$ and $\eta$ from Eq.~(\ref{eq:source_spec})}
As a free parameter, we first add a universal source slope~$\alpha$.
We then allow $\eta$, parameterising a universal low-energy shape of all spectra,
to be a second free parameter.
In addition to B/C constraining the transport parameters, some primary species must
be added to constrain $\alpha$ and $\eta$. We restrict ourselves to O,
the most abundant boron progenitor, because it was measured by both the HEAO-3
experiment~\citep{1990A&A...233...96E}, and also the TRACER
experiment~\citep{2008ApJ...678..262A}. 
The modulation levels were $\Phi=250$~MV for HEAO-3 and $\Phi=500$~MV for TRACER.
We estmated the latter number from the solar activity at the time of flight
(2 weeks in December 2003) as seen from neutron monitors
data\footnote{{\tiny http://ulysses.sr.unh.edu/NeutronMonitor/Misc/neutron2.html}.
Indeed, the solar activity between 2002 and 2004 has not varied much, so that we use a value
for $\Phi$ derived from the BESS 2002 flight (see Fig.~2 of~\citealt{2007APh....28..154S}).}.

\begin{figure}[t!]
\centering
\includegraphics[width = .5\textwidth]{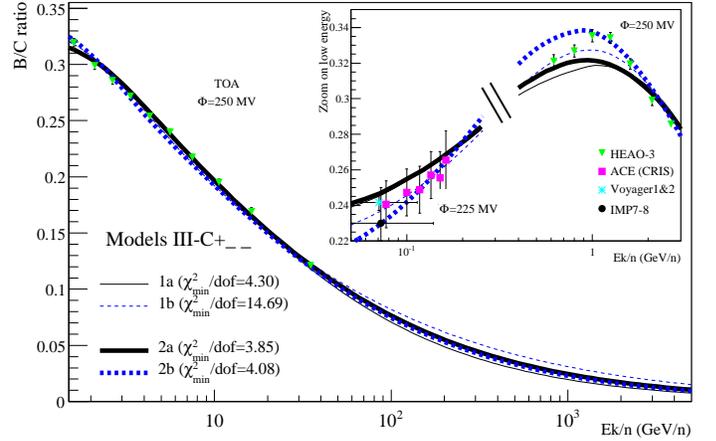}
\caption{B/C ratio from best-fit models of Table~\ref{tab:resultsBestFitIIIC+}.
In addition to the propagation parameters, free parameters of the source spectra
are $\alpha$ (thin lines, labeled 1) or $\alpha$ and $\eta$ (thick lines, labeled 2):
The two models are tested on two datasets for the primary flux (in addition to
using the B/C HEAO-3 data): O is as measured by HEAO-3 (black lines, dataset a) or
as measured by TRACER (blue lines, dataset b).
}%
\label{fig:IIIC+1and2B/C}
\end{figure}

In total, we test four models (denoted by 1a, 1b, 2a, and 2b for legibility): 
\begin{itemize}
  \item III-C+1a: $\{\lambda_0,\, R_0,\, \delta,\, {\cal V}_a\} + \{\alpha\}$, with O=HEAO-3;
	\item III-C+1b:  $\{\lambda_0,\, R_0,\, \delta,\, {\cal V}_a\} + \{\alpha\}$, with O=TRACER;
	\item III-C+2a: $\{\lambda_0,\, R_0,\, \delta,\, {\cal V}_a\} + \{\alpha,\, \eta\}$, with O=HEAO-3;
	\item III-C+2b:  $\{\lambda_0,\, R_0,\, \delta,\, {\cal V}_a\} + \{\alpha,\, \eta\}$, with O=TRACER;
\end{itemize}
where the Arabic numbers relate to the source-spectrum free parameters used in the calculation,
and the lower case relates to the chosen oxygen-flux dataset (a=HEAO-3, b=TRACER).
The most probable parameters are gathered in Table~\ref{tab:IIIC+1and2}, where,
to provide a comparison, the first line reports the values found for Model~III-C
(i.e. with $\gamma\equiv\alpha+\delta$ fixed to 2.65). 
\begin{table*}[t]
\centering
\begin{tabular}{cccccccc} \hline\hline
Model-Data & $\lambda_{0}$   &      $R_0$          &         $\delta$        &  ${\cal V}_a$ & $\alpha$ & $\eta$ &  $10^{20}\times(q_C|q_N|q_O)^\dagger$ \\ 
      &    g cm$^{-2}$  &       GV            &                         & km s$^{-1}$kpc$^{-1}$  & & & (m$^3$~s~GeV/n)$^{-1}$\\\hline
& \multicolumn{7}{c}{} \\ 
 III-C$^\ddagger$     & $27^{+2}_{-2}$ & $2.6^{+0.4}_{-0.7}$ & $0.53^{+0.02}_{-0.03}$ & $86^{+9}_{-5}$ &            -              & - &  -  \vspace{0.3cm}\\
III-C+1a   & $37^{+2}_{-2}$ & $4.4^{+0.1}_{-0.2}$ & $0.61^{+0.01}_{-0.01}$ & $64^{+4}_{-4}$ & $2.124^{+0.005}_{-0.007}$ & - &  - \vspace{0.1cm}\\
III-C+1b   & $20.9^{+0.2}_{-0.8}$ & $0.3^{+0.6}_{-0.1}$ & $0.47^{+0.01}_{-0.01}$ & $103^{+2}_{-3}$ & $2.294^{+0.004}_{-0.006}$ & - &  - \vspace{0.1cm}\\
III-C+2a   & $29^{+2}_{-2}$ & $2.7^{+0.3}_{-0.4}$ & $0.55^{+0.01}_{-0.02}$ & $84^{+4}_{-7}$ & $2.16^{+0.01}_{-0.01}$ & $0.3^{+0.1}_{-0.2}$ &  - \vspace{0.1cm}\\
III-C+2b   & $32^{+4}_{-1}$ & $4.3^{+0.3}_{-0.1}$ & $0.56^{+0.03}_{-0.01}$ & $62^{+2}_{-2}$ & $2.14^{+0.03}_{-0.01}$ & $-6.7^{+0.9}_{-0.1}$ &  - \vspace{0.3cm}\\
III-C+4a   & $40^{+3}_{-1}$ & $4.6^{+0.2}_{-0.1}$ & $0.64^{+0.01}_{-0.02}$ & $58^{+2}_{-5}$ & $2.13^{+0.01}_{-0.01}$ & - &
$\;\;\;1.93^{+0.04}_{-0.004}|0.089^{+0.007}_{-0.005}|2.42^{+0.04}_{-0.05}$ \vspace{0.1cm}\\
III-C+5a   & $38^{+1}_{-2}$ & $4.4^{+0.1}_{-0.3}$ & $0.60^{+0.02}_{-0.01}$ & $81^{+4}_{-1}$ & $2.17^{+0.02}_{-0.02}$ & $-0.4^{+1.2}_{-0.1}$ &
$2.2^{+0.2}_{-0.1}|0.107^{+0.01}_{-0.006}|2.7^{+0.3}_{-0.1}$ \vspace{0.1cm}\\\hline
\end{tabular}
{\\\scriptsize $^\ddagger$ III-C: propagation parameters are $\{\lambda_0,\, R_0,\, \delta,\, {\cal V}_a\}$ and the B/C dataset is HEAO-3+Voyager$\,$1$\,$\&$\,$2+IMP7-8.\\
$^\dagger$ Abundances are $1.65|0.10|2.04$ for HEAO-3 (\citealt{1990A&A...233...96E}, and see Table~\ref{tab:abundances}).}
\caption{Most probable values of the propagation parameters 
(after marginalising over the other parameters) for models III-C+\_\,\_\,. The additional
free parameters and data ''\_\,\_\," correspond to:
1a/1b\,=$\{\alpha\}$, 2a/2b=$\{\alpha,\, \eta\}$, 4a/4b\,=$\{\alpha,\,q_C,\,qN,\,qO\}$, 
5a/5b=$\{\alpha,\, \eta,\,q_C,\,qN,\,qO\}$ with either O data is HEAO-3 (a) or TRACER (b). 
The uncertainty on the parameters correspond to 68\% CL of the
marginalised PDF (see App.~\ref{s:CL}). The associated best-fit parameters are gathered in
Table~\ref{tab:resultsBestFitIIIC+}.}
\label{tab:IIIC+1and2}
\end{table*}
We remark that by adding HEAO-3 oxygen data to the fit (1a), the propagation parameters $\lambda_0$, $R_0$,
and $\delta$ overshoot Model~III-C's results, while they undershoot those of Model 1b (TRACER data). 
The parameter ${\cal V}_a$ undershoots and overshoots for these two
models respectively, since it is anti-correlated with the former parameters. 
As a conse\-quence, the fit to B/C is worsened, especially at low energy
(see Fig.~\ref{fig:IIIC+1and2B/C}).

\begin{figure}[t!]
\centering
\includegraphics[width = .24\textwidth]{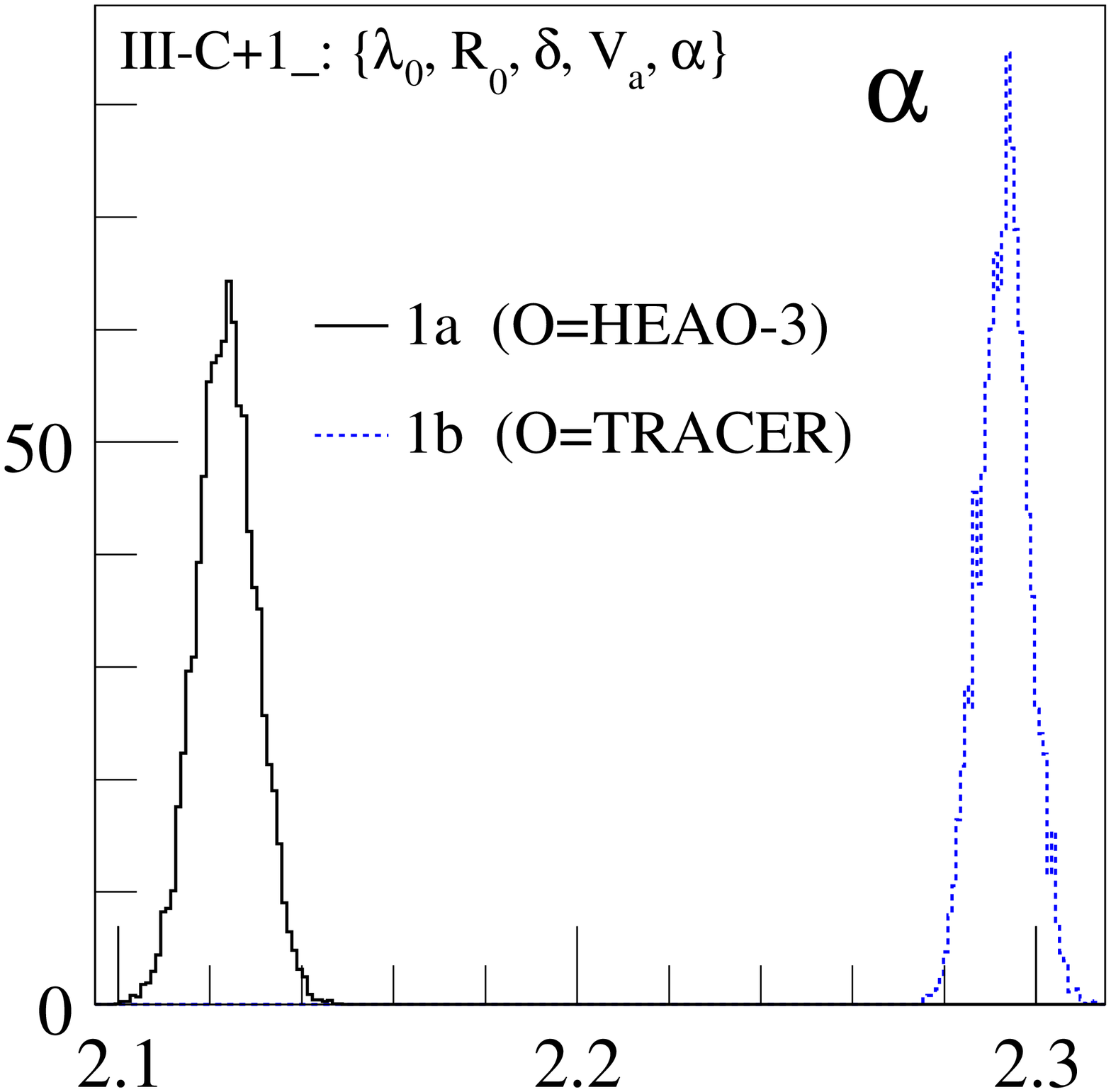}
\includegraphics[width = .24\textwidth]{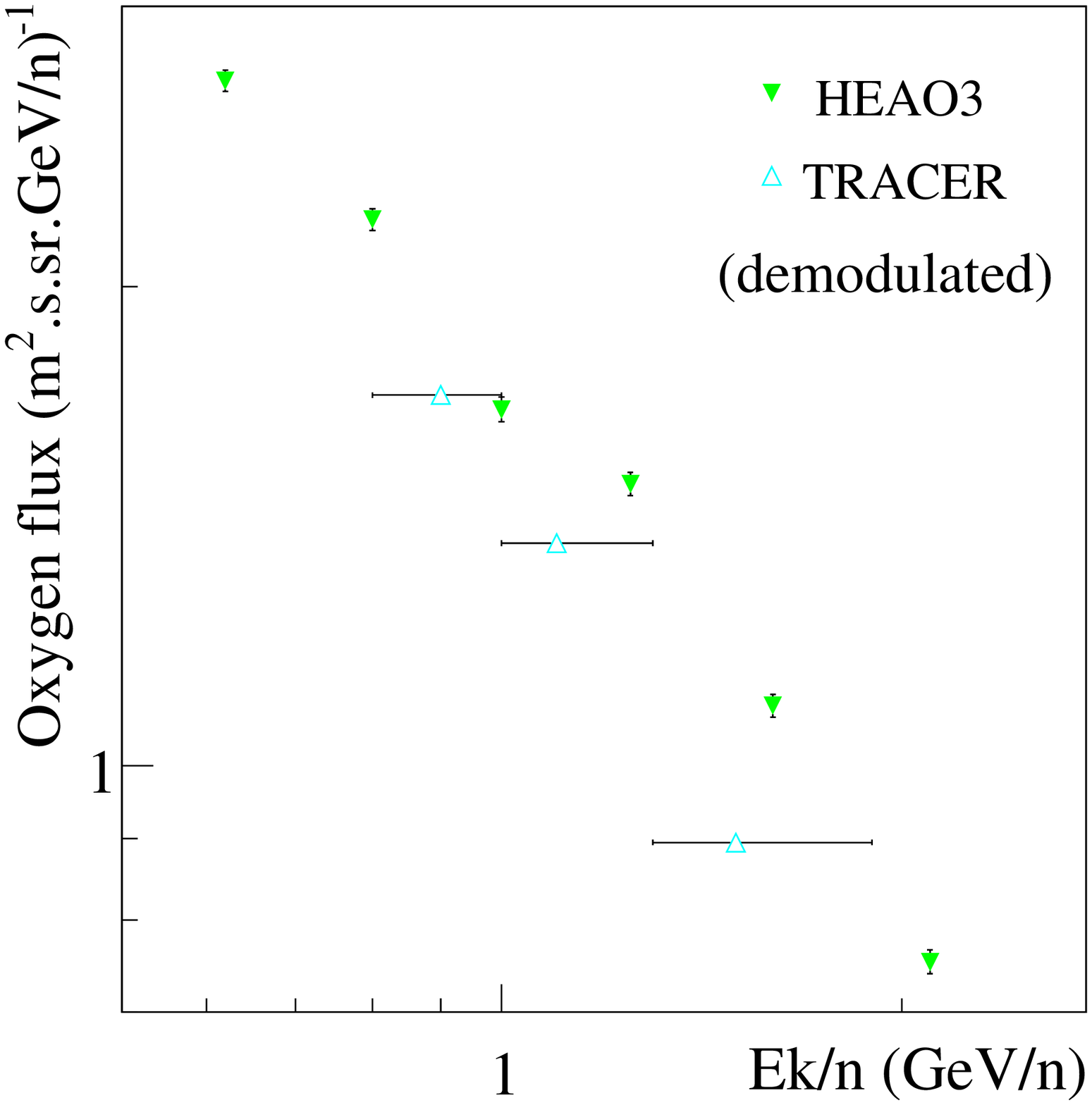}
\includegraphics[width = .5\textwidth]{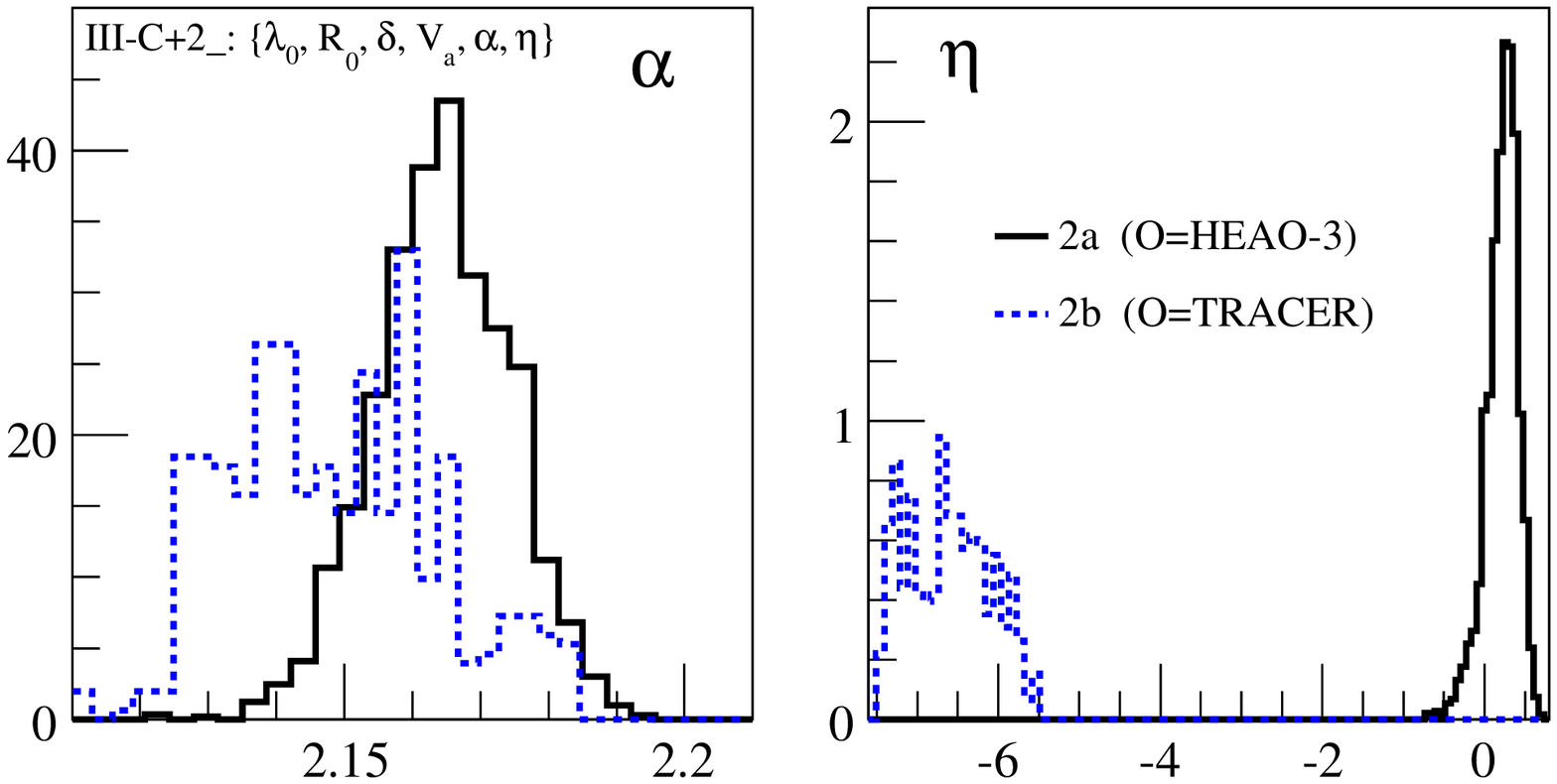}
\caption{Marginalised PDF for models~III-C+1 ($\alpha$ free parameter, top left)
and III-C+2 ($\alpha$ and $\eta$ free parameters, bottom panels). In the three panels,
the solid-black lines rely on HEAO-3 oxygen data, whereas the blue-dashed lines
rely on TRACER oxygen data (thin and thick lines are as in Fig.~\ref{fig:IIIC+1and2B/C}).
Top right panel: zoom in demodulated O low energy HEAO-3
and TRACER data. (The solid segments on TRACER data show the energy bin size.
Uncertainty on all fluxes are present, but too small to notice).}
\label{fig:alphaIIIC+1}
\end{figure}

\begin{figure}[t!]
\centering
\includegraphics[width = .5\textwidth]{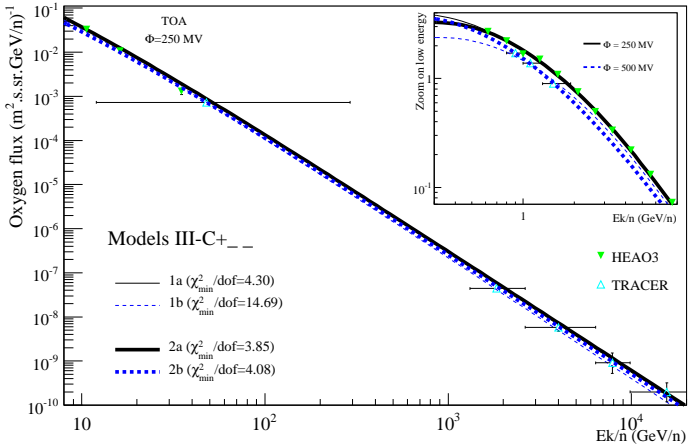}
\caption{Same models as in Fig.~\ref{fig:IIIC+1and2B/C}, but for the oxygen flux.}%
\label{fig:IIIC+1and2O}
\end{figure}

The top left panel of Fig.~\ref{fig:alphaIIIC+1} shows the slopes $\alpha$ derived for
Models 1a (solid black) and 1b (dashed blue). In both cases, $\alpha$
is well constrained, but the values are inconsistent, a result that is clear because
the low-energy data are also inconsistent: the demodulated (i.e. IS) HEAO-3 and TRACER
oxygen data points are shown in the right panel of Fig.~\ref{fig:alphaIIIC+1}.
To remedy this situation, we allow $\eta$ to be a free parameter (family of models~III+2).
The net effect is to absorb any uncertainty originating in either the modulation level
or the source-spectrum low-energy shape. As shown in the bottom panel of Fig.~\ref{fig:alphaIIIC+1},
the source slopes derived from the two experiments are now in far closer agreement (bottom left),
with $\alpha\simeq 2.15$. The most probable values and the best-fit model values are
given in Tables~\ref{tab:IIIC+1and2} and \ref{tab:resultsBestFitIIIC+} respectively. The
effect of this action is evident in the low-energy slope of the source spectrum $\eta$.
As seen in the bottom-right
panel, the two data sets contain significantly inconsistent ranges.
The value $\eta_{\rm TRACER}\simeq -6.7$ probably indicates that
the solar modulation we chose was incorrect. The value $\eta_{\rm HEAO-3}\simeq 0.3$
might provide a reasonable guess of the low-energy shape of the source spectrum, but
might also be a consequence of systematics in the experiment. The associated
oxygen fluxes are shown in Fig.~\ref{fig:IIIC+1and2O} for the best-fit models:
as explained, models that allow $\eta$ to vary (thick lines) reproduce more accurately
the data than when $\eta$ is set to be -1 (thin lines).

Although it would be precipitate to draw any firm conclusion about the low-energy shape, we can turn the argument
around to serve as a diagnosis of the low-energy data quality. For instance,
assuming that the spectral index $\alpha$ of all elements was the same, extracting and comparing
$\eta_i$ for each of these $i$ elements may enable us to infer some systematics
remaining in the data. It would be worth fitting the H and He species, which are the 
most reliably measured fluxes to date; this will be considered in a future study
using diffusion models.

	\subsubsection{$\alpha$, $\eta$ and source normalisation $q_i$\label{sec:fit_final}}

The final two models add, as free parameters, the CNO element source abundances 
(relative isotopic abundances are fixed to SS ones). The data used in the fit
are B/C, C, N, and O, all measured by HEAO-3 (TRACER data for C and N have not yet
been published). The models, which are denoted by short 4a and 5a in the text below,
are:
\begin{itemize}
  \item III-C+4a: $\{\lambda_0,\, R_0,\, \delta,\, {\cal V}_a\} + \{\alpha, q_C,\, q_N,\, q_O\}$;
	\item III-C+5a: $\{\lambda_0,\, R_0,\, \delta,\, {\cal V}_a\} + \{\alpha,\, \eta, q_C,\, q_N,\, q_O\}$.
\end{itemize}

\begin{figure*}[t!]
\centering
\includegraphics[width = \textwidth]{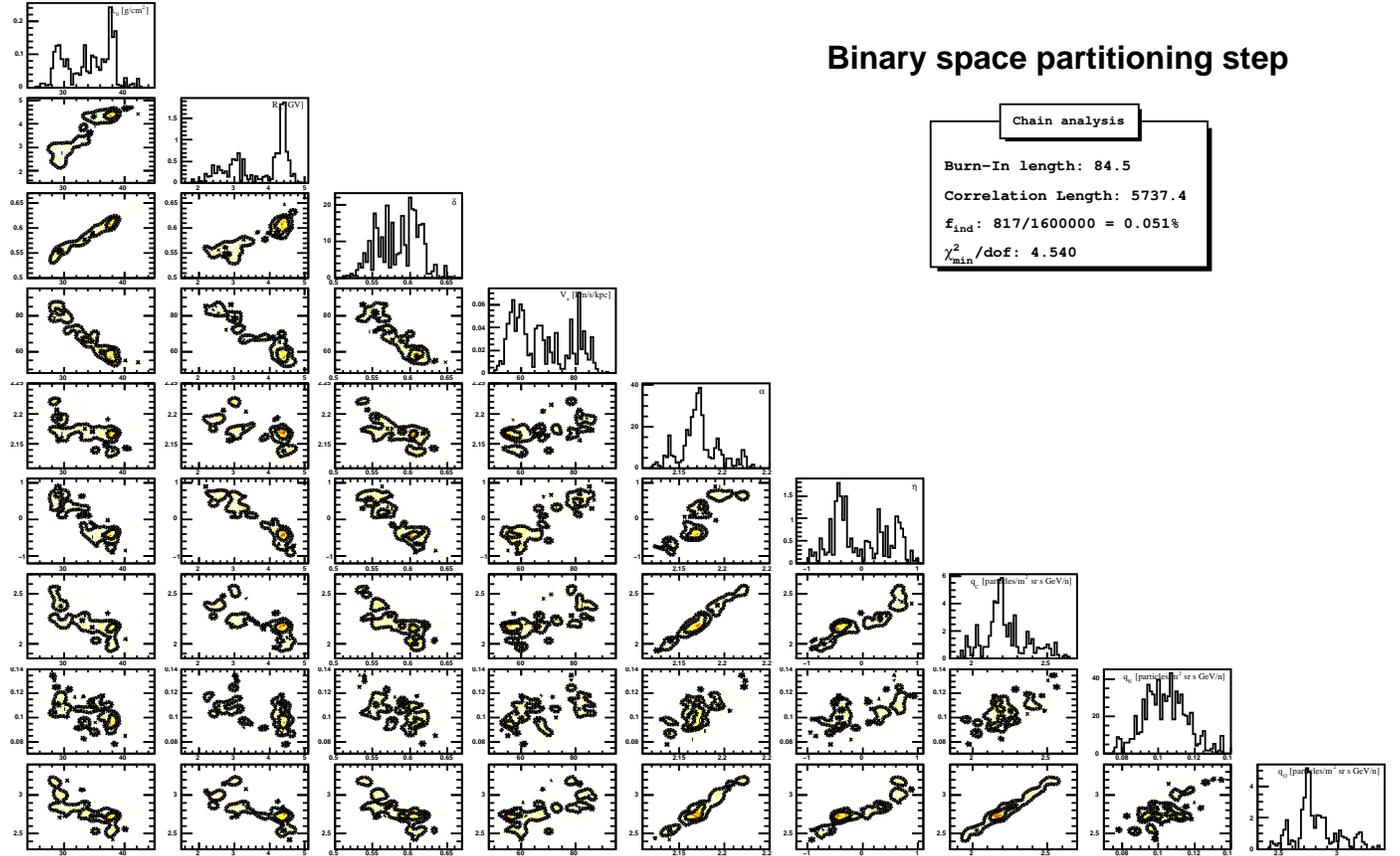}
\caption{PDF (diagonal) and 2D correlations (off-diagonal) plots 
for the nine free parameters of Model~III-C+5 when fitted on B/C, and CNO HEAO-3 data.}
\label{fig:correl9params}
\end{figure*}
The PDF and 2D correlations plots between propagation and source parameters
are seen in Fig.~\ref{fig:correl9params}.
With nine parameters, the efficiency is very low ($f_{\rm ind}\lesssim 0.05\%$),
even using the BSP trial function. To obtain $\sim 800$ independent points,
a total $1.6\cdot 10^6$ steps were completed. The contours are not as regular as for our 4-parameter
model (see Fig.~\ref{fig:Model_IIandIII}), but correlations between the parameters
are still clearly evident: we recover the $\lambda_0-R_0-\delta$ correlations
(and anti-correlation with ${\cal V}_a$). In addition, we note that
all source-related parameters ($\alpha$, $\eta$ and element abundances
$q_{C,N,O}$) are correlated among themselves and with ${\cal V}_a$ (hence anti-correlated
with the remaining transport parameters). This is especially visible for the
primary species C and O, while less clear for the mixed species N.
This is understood as the parameter $\gamma$, i.e. the slope of the propagated
primary fluxes ($\gamma=\alpha+\delta$) is mostly fixed by measurements: if we decrease
$\delta$, then $\alpha$ must be
increased to match the data. However, if the source slope is increased,
the exponent $\eta$ must also increase to match the low-energy data.
The positive correlation between source abundances comes from the
fact that relative fluxes should be preserved.

The most probable values are gathered in Table~\ref{tab:IIIC+1and2}.
Compared with the respective Models~1a and 2a, leaving
the source abundances $q_C$, $q_N$ and $q_O$ free in 4a and 5a
does not significantly change our conclusions. Again, adding $\eta$
(2a and 5a) as a free parameter allows us to absorb the
low-energy uncertainties in the data, so that we obtain $\alpha=2.17$ 
(5a) instead of the biased value of $2.13$ (4a). The same conclusions
hold for other propagation parameters. On the derived source 
abundances, the impact of adding the parameter $\eta$ is for them to increase.
The relative C:N:O abundances (O$\equiv 1$)
are respectively $0.78:0.36:1$ (4a) and $0.82:0.40:1$ (5a), the second
model providing values slightly closer to those derived from HEAO-3 data $0.81:0.49:1$.

The difference in the source element abundances when they are rescaled to match the
data or including them in the MCMC is also seen from Table~\ref{tab:resultsBestFitIIIC+},
which gathers the best-fit parameters. The next-to-last line reproduces $q_{C,N,O}$ obtained
for all models: all abundances are roughly in agreement, although our approach
underlines the importance of taking the correlations between the parameters
properly into account in extracting unbiased estimates of the propagation and
source parameters.
\begin{table*}[t!]
\centering
\begin{tabular}{cccccccccc} \hline\hline
Model-Data & $\lambda_{0}^{\rm best}$ & $R_0^{\rm best}$ & $\delta^{\rm best}$ & ${\cal V}_a^{\rm best}$ & $\alpha^{\rm best}$
& $\eta^{\rm best}$ & $10^{20}\times(q_C|q_N|q_O)^\dagger$ & $\chi^2$/dof   \\
      &    g cm$^{-2}$  &       GV            &                       & km s$^{-1}$kpc$^{-1}$  & & & (m$^3$~s~GeV/n)$^{-1}$ & \\\hline
& \multicolumn{7}{c}{} \\ 
III-C    & 26.9 & 2.45 &  0.527 & 88.5 & - & - & ~~$[1.48|0.08|1.85]^\ddagger$ & 1.06 \vspace{0.1cm}\\
III-C+1a & 36.9 & 4.34 &  0.610 & 64.6 & 2.123 & - & ~~$[1.92|0.105|2.40]^\ddagger$ & 4.30 \vspace{0.1cm}\\
III-C+1b & 20.7 & 0.46 &  0.470 & 102.9 & 2.293 & - & ~~$[3.43|0.219|4.12]^\ddagger$ & 14.69 \vspace{0.1cm}\\
III-C+2a & 28.7 & 2.61 &  0.547 & 84.5 & 2.168 & 0.305 & ~~$[2.25|0.126|2.81]^\ddagger$ & 3.85 \vspace{0.1cm}\\
III-C+2b & 33.0 & 4.24 &  0.568 & 61.5 & 2.154 & -6.545 & ~~$[2.09|0.161|2.17]^\ddagger$ & 4.08 \vspace{0.3cm}\\
III-C+4a & 39.2 & 4.60 &  0.626 & 59.2 & 2.126 & - & $1.92|0.090|2.42$ & 4.65 \vspace{0.1cm}\\
III-C+5a & 28.6 & 2.44 &  0.545 & 83.0 & 2.175 & 0.449 & $2.27|0.104|2.86$ & 4.54 \vspace{0.1cm}\\
\hline
\end{tabular}
{\\\scriptsize $^\ddagger$ These values are not extracted from the PDF: they
are values of rescaled abundances required to match HEAO-3 CNO data at 10.6 GeV/n.\\
$^\dagger$ For a comparison, HEAO-3 abundances (\citealt{1990A&A...233...96E}, and see Table~\ref{tab:abundances}) are $1.65|0.10|2.04$.}
\caption{Best-fit values (corresponding to $\chi^2_{\rm min}$)
for all models as given in Table~\ref{tab:IIIC+1and2}. Number of data points for
the $\chi^2_{\rm min}/$dof calculation: 22 B/C data (III-C=HEAO-3+Voyager$\,$1$\,$\&$\,$2+IMP7-8)
plus 14 oxygen HEAO-3 data for 1a and 2a, 8 oxygen TRACER data for 1b and 2b,
or $14\times3$ (C, N and O) HEAO-3 data for 4a and 5a.}
\label{tab:resultsBestFitIIIC+}
\end{table*}

\begin{figure}[t!]
\centering
\includegraphics[width = 0.15\textwidth]{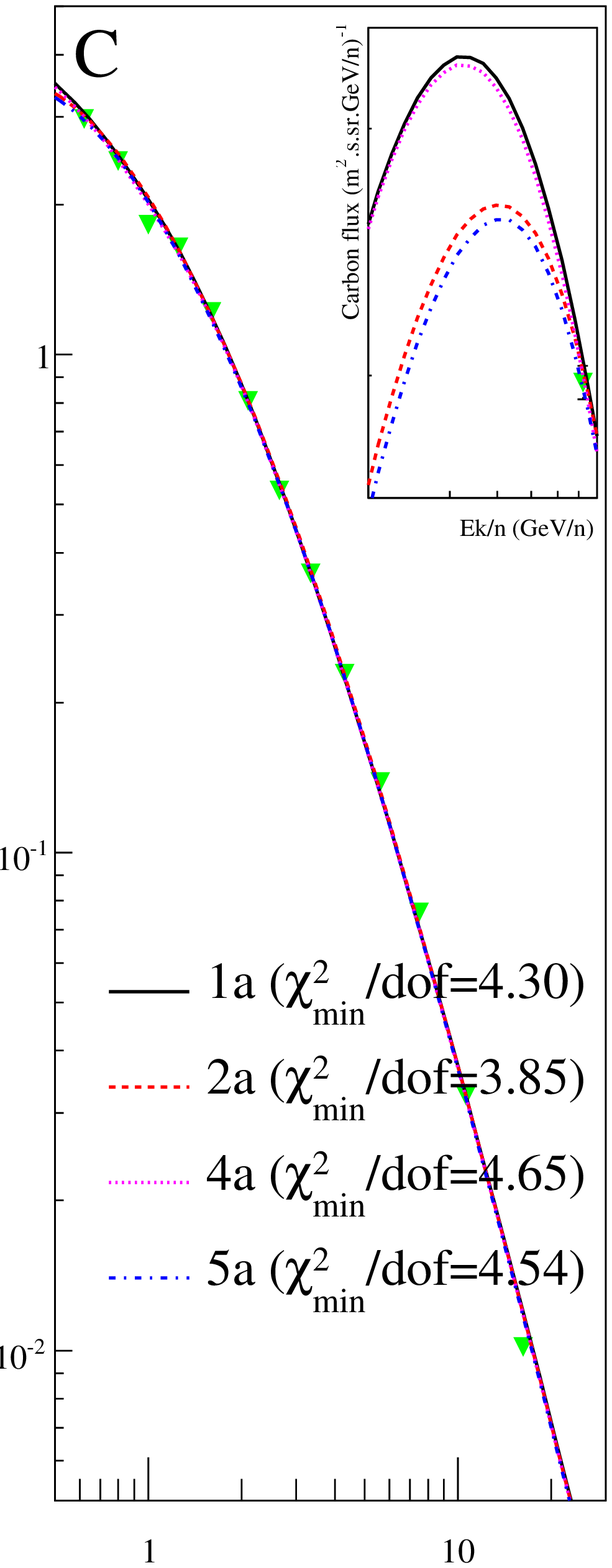}
\includegraphics[width = 0.15\textwidth]{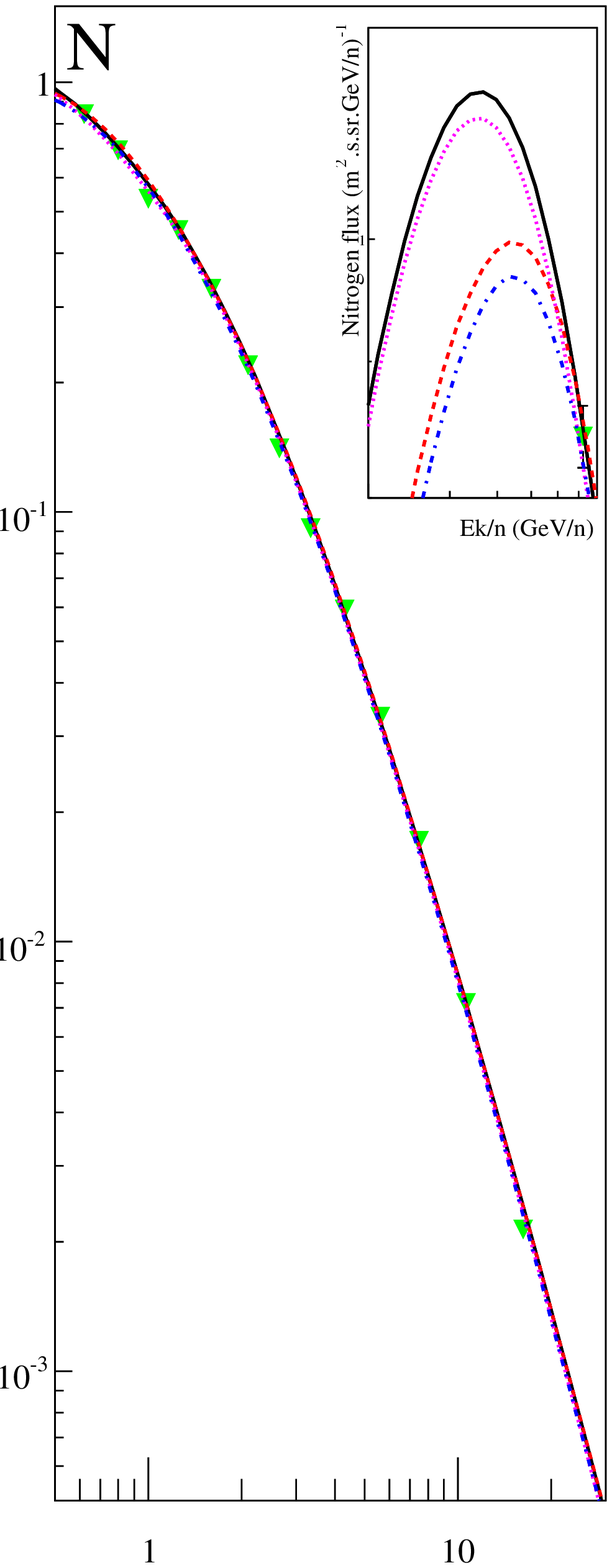}
\includegraphics[width = 0.15\textwidth]{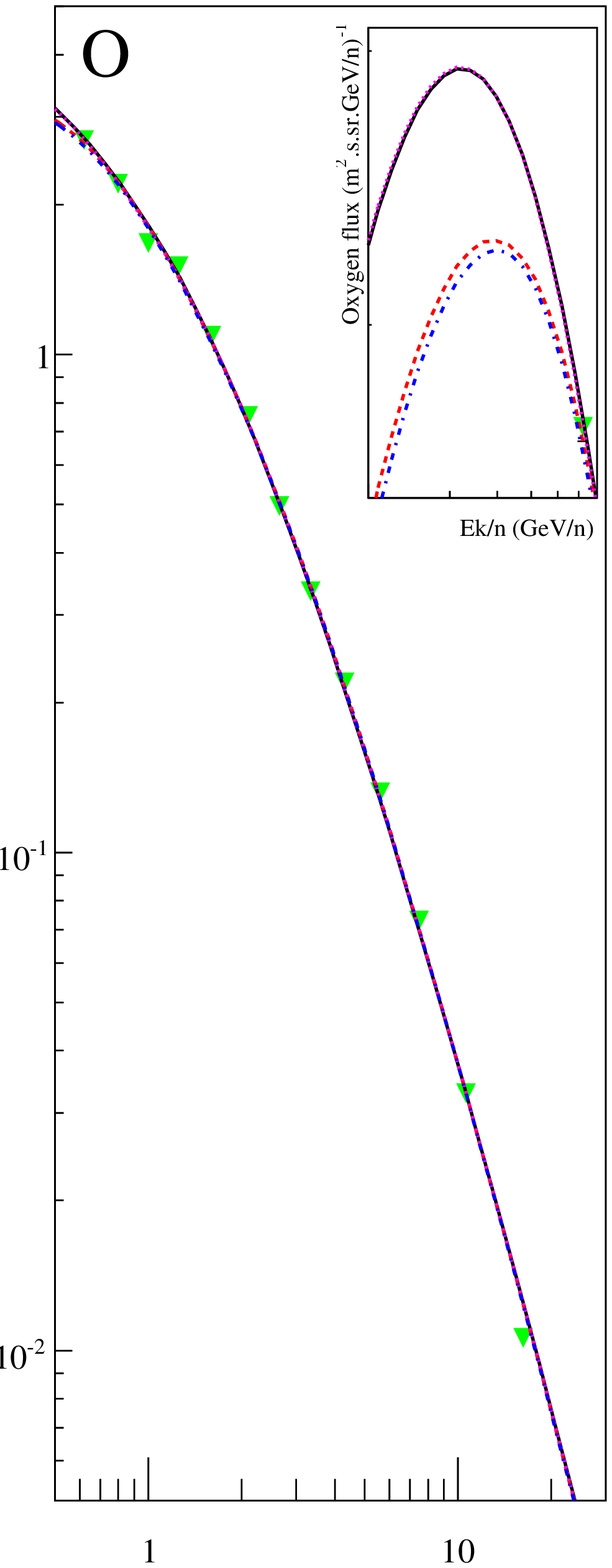}
\caption{Carbon, nitrogen and oxygen fluxes from best-fit models of
Table~\ref{tab:resultsBestFitIIIC+}.}%
\label{fig:IIIC+1and2CNO}
\end{figure}

The goodness of fit for the models when applied to the B/C, C, N, and O data is shown in
the last column of Table~\ref{tab:resultsBestFitIIIC+}, in terms of the $\chi^2_{\rm min}$
value. The models in which $q_{C,N,O}$ is free do not provide a closer match between
models and data but also a no poorer fit than when $q_{C,N,O}$ is fixed.
As soon as primary fluxes are included in the fit
(compared to Model~III-C), the $\chi^2_{\rm min}$ is worsened.
This is due to a combination of an imperfect fit to the primary fluxes and,
as already said, a poorer B/C fit because the propagation parameters are
optimised to match the former rather than the latter (B/C). The best-fit parameters are given in the
same Table, and the associated CNO fluxes are plotted in Fig.~\ref{fig:IIIC+1and2CNO}
for illustration purposes.

\subsubsection{Perspective on source spectrum parameters}

Other primary species could have been included in the $\chi^2$ calculation to
i) constrain further $\alpha$, and/or ii) to check the hypothesis $\alpha_i\neq\alpha_j$
for different species, and/or iii) diagnose some problems in the data, if we believe
the slope should be universal.  
However, using a few primary species (O or CNO) already affects
the goodness of fit of B/C (compare model III-C to others in Fig.~\ref{fig:IIIC+1and2B/C}).
Since there are many more measured primary fluxes than secondary
ones, taking too many primaries would weigh too significantly in the $\chi^2$,
compared to the B/C contribution, and this would divert the MCMC into regions of
the parameter space that fit these fluxes rather than B/C. Since systematic errors
are known to be larger in flux
measurements than in ratios, this may lead to biased estimates of the propagation parameters.
Allowing $\eta$ to be a free parameter is a first step to decrease the impact
of this bias (in the low-energy part of the spectrum).
These biases are not necessarily an issue since we may be more
interested in estimates of the source parameters rather than in unbiased
value of the propagation parameters.

To illustrate the difficulty in data systematics, it suffices to say that for the most
abundant species H and He, all experiments provided mutually incompatible
measurements, until AMS and BESS experiments flew ten years ago. We cannot therefore
expect HEAO-3 data, acquired in 1979, to be completely free of such drawbacks. Again, we await
the publication of several forthcoming new data sets before pursuing our analysis further 
in this direction.

\section{Conclusion}
\label{s:conclusions}

We have implemented a Markov Chain Monte Carlo to extract the posterior distribution functions
of the propagation parameters in a LBM. Three trial functions were used, namely
a standard Gaussian step, an N-dimensional Gaussian step and its covariance matrix, and a
binary-space partitioning.
For each method, a large number of chains were processed in parallel to accelerate the PDF calculations.
The three trial functions were used sequentially, each method providing some inputs to the next:
while the first one was good at identifying the general range of the propagation parameters, it
was not as efficient in providing an accurate description of the PDF. The two other methods
provided this accuracy, and the final results were based on PDF obtained from the chains
processed with the binary-space partitioning.

Taking advantage of the sound statistical properties of the MCMC, confidence intervals for the
propagation parameters can be given, as well as confidence contours for all fluxes
and other quantities derived from the propagation parameters. The MCMC was also used
to compare the impact of choosing different datasets and ascertain the merits of
different hypotheses concerning the propagation models.
Concerning the first aspect, we have shown that combining different B/C datasets leaves
mostly unchanged the propagation parameters, while strongly affecting the assessment
of the goodness of a model. We also show that at present, the \pbar\ data do not cover
a sufficiently large energy range to constrain the propagation parameters, but they
could be useful in crosschecking the secondary nature of the flux in the future.

In this first paper, we have focused on the phenomenologically well-understood LBM,
to ease and simplify the discussion and implementation of the MCMC. In agreement with
previous studies, we confirm that a model with a rigidity cutoff performs more successfully
than one without and that reacceleration is preferred over no reacceleration. Such a model
can be associated with a diffusion model with wind and reacceleration. As found in
\citet{2001ApJ...555..585M}, the best-fit models demand both a rigidity cutoff (wind)
and reacceleration, but do not allow us to reconcile the diffusion slope
with a Kolmogorov spectrum for turbulence. An alternative
model with two slopes for the diffusion was used, but it is not favoured by the data.
In a last stage, we allowed the abundance and slope of the source spectra to be free
parameters, as well as the element abundances of C, N, and O. This illustrated a
correlation between the propagation and source parameters, potentially biasing the
estimates of these parameters. The best-fit model slope for the source abundances 
was $\alpha\approx 2.17$ using HEAO-3 data, compatible with
the value $\alpha\approx 2.14$ for TRACER data. The MCMC approach allowed us
to draw confidence intervals for the propagation parameters, the source parameters,
and also for all fluxes.

A wealth of new data on Galactic CR fluxes are expected soon. As illustrated for the LBM,
the MCMC is a robust tool in handling complex data and model parameters, 
where one has to fit simultaneously all source and propagation parameters.
The next step is to apply this approach to more realistic diffusion models
and larger datasets, on a wider range of nuclear species.

\begin{acknowledgements}
D.M. and R.T. warmly thank Joanna Dunkley for useful discussions 
at an early stage of this work. We thank C\'eline Combet for a careful
reading of the paper.
\end{acknowledgements}

\begin{appendix}

   \section{Best fit, goodness of a model, most probable values and confidence levels/intervals}
   \label{s:CL}
The best-fit model parameters are given by a unique set of parameters for which the
$\chi^2$ value of Eq.~(\ref{eq:chi2}) is minimized;
the goodness of fit of a model is given by $\chi^2_{\rm min}/$dof.
On the other hand, the most probable value for each parameter $\theta_i$ is defined as the maximum 
${\cal P}^{\rm max}_i\equiv {\cal P}(\theta_i^{\rm max})$ of its PDF (after marginalising).
The most probable $\vectheta^{\rm max}$ and best-fit model parameters $\vectheta^{\rm best}$
do not necessarily coincide, especially when correlations exist between parameters.
The best-fit model parameters are best suited to providing the most likely
CR fluxes, whereas the 1D marginalised PDF provides directly the most likely value
of the parameter.

\subsection{Confidence levels/intervals on parameters\label{s:CLparam}}

Confidence intervals (CI), associated with a confidence level (CL), are
constructed  from the PDF.
The asymmetric interval $\Delta_x\equiv [\theta_i^{\rm max}-\theta^-_x,
\theta_i^{\rm max}+\theta^+_x]$ such as
\begin{equation}
{\rm CL}(x)\equiv \int_{\Delta_x}{\cal P}(\theta_i) {\rm d} \theta_i= 1 - \gamma,
\end{equation}
defines the $1 - \gamma$ confidence level (CL), along with the CI of the
parameter $\theta_i$. Here, the CIs (i.e $\theta^-_x$ and $\theta^+_x$)
are found by decreasing the value ${\cal P}(\theta_i)$ from
${\cal P}^{\rm max}_i$ to ${\cal P}_i^x$, such that $1-\gamma=x$.
This is easily generalised to 2D confidence levels in constructing
2D confidence intervals as shown later in correlation plots. Below, we use the
$x=68\%$ and $x=95\%$ CLs, corresponding to $1\sigma$ and $2\sigma$
uncertainties.

\subsection{Confidence intervals on fluxes\label{s:CLflux}}
The best-fit model fluxes (e.g., B/C, O, $\bar{p}$) are calculated
from the best-fit model parameters. Confidence levels in these quantities
cannot be obtained from the 1D marginalised CIs of the 
parameters. They must be constructed from
a sampling of the (still) correlated parameters. This is achieved
by using all sets of parameters $\{\vectheta\}_{x\% \rm CL}=\{\vectheta_i\}_{i=1\cdots p}$,
for which $\chi^2(\vectheta_i)$ falls in the $x\%$ confidence level of the $\chi^2$ PDF.
Once these sets are found, we simply calculate the desired flux for all
the sets: the maximum and minimum values are kept for each energy bin, defining
confidence envelopes for this flux. Thus, the main task is to construct
confidence intervals for the $\chi^2$ distribution. 

For $n$ parameters in the large sample limit|where the joint PDF for
the estimator of the parameters and the likelihood function become Gaussian|,
the CI is given by 
$$
[\chi^2_{\mathrm{min}},\, \chi^2_{\mathrm{min}}+\Delta \chi^2], \quad
{\rm where} \quad \Delta \chi^2 = Q_{\gamma} (1 - \gamma, n)
$$
is the quantile of order $1-\gamma$ (confidence level CL) of the $\chi^2$
distribution \citep{1997sda..book.....C}. 
However, by applying the MCMC, we have access to a direct sampling of the
$\chi^2$ distribution. Hence, independently of the statistical meaning of a model,
the confidence interval is extracted from the cumulative  $\chi^2$ PDF,
by requiring that
\begin{equation}
\int_{\chi^2_{\rm min}}^{\chi^2_{\rm min}+\Delta \chi^2}{\cal P}(\chi^2) {\rm d} \chi^2 =1 - \gamma.
\label{eq:CI_PDF}
\end{equation}
We nevertheless checked that both approaches provide very similar results.
For instance, the CIs (for Model~III-C) obtained directly from
Fig.~\ref{fig:chi2_envelopes} are ${\rm CI}~(68\%)=[\chi^2_{\rm min}, \chi^2_{\rm min}+4.9]$
and ${\rm CI}~(95\%)=[\chi^2_{\rm min}, \chi^2_{\rm min}+9.2]$, whereas
they are ${\rm CI}~(68\%)=[\chi^2_{\rm min}, \chi^2_{\rm min}+4.7]$ and
${\rm CI}~(95\%)=[\chi^2_{\rm min}, \chi^2_{\rm min}+9.5]$ when calculated
from the $Q_{\gamma} (1 - \gamma, n)$ quantiles \citep{1997sda..book.....C}.

\section{Data\label{s:config_data}}
In the paper, we focus on the B/C ratio, which is the most accurate
tracer of the propagation parameters (other tracers, such
as the sub-Fe/Fe or the quartet $^1$H, $^2$H, $^3$He and $^4$He
are not considered). We also estimate the potential
of the \pbar, a secondary species, as an alternative species
for constraining these parameters. We describe below the typical
configurations used to calculate the corresponding spectrum
as well as the associated datasets used.

	\subsection{B/C\label{s:config_dataBC}}
The default configuration for nuclei is the following:
the value of the observed propagated slope $\gamma=\alpha+\delta$,
unless stated otherwise,
is set to be 2.65 \citep{2008ApJ...678..262A}, and source abundances
of the most abundant species (C, N, O, F, Ne, Na, Mg, Al, Si) are
rescaled to match HEAO-3 data at 10.6 GeV/n. Boron is assumed to be a
pure secondary species. Only elements lighter than Si are propagated,
since they are the only relevant ones for determining B/C \citep{2001ApJ...555..585M}.

For B/C at intermediate GeV energies, we use HEAO-3 data
\citep{1990A&A...233...96E}. They are complemented at low energy
by the ACE (CRIS) data \citep{2006AdSpR..38.1558D}.
For a few model iterations, we also look for combined constraints
from multiple sets of data. A collection of low-energy data is formed
by data sets for the IMP7-8~\citep{1987ApJS...64..269G}, ISEE-3~\citep{1988ApJ...328..940K},
Ulysses~\citep{1996A&A...316..555D}, and Voyager~1\&2 \citep{1999ICRC....3...41L}
spacecrafts. At higher energy, we consider several
balloon flights \citep{1978ApJ...223..676L,1978ApJ...226.1147O,1980ApJ...239..712S,
1987ApJ...322..981D}, the ATIC-2 balloon-borne experiment \citep{2007arXiv0707.4415P},
and the Spacelab-2 experiment \citep{1991ApJ...374..356M}.
For element fluxes, HEA0-3 and TRACER results \citep{2008ApJ...678..262A} are used.

	\subsection{$\bar{p}$\label{s:config_dataPbar}}

For the calculation of the \pbar\ flux, the situation is simpler:
the production of this secondary flux can be directly linked to the accurate
measurements of propagated p and He fluxes, by the
AMS \citep{2000PhLB..472..215A,2000PhLB..490...27A,2000PhLB..494..193A}
and BESS experiments \citep{2000ApJ...545.1135S,2007APh....28..154S}.
For more details of the \pbar\ flux calculation (cross sections,
source terms, \dots), the reader is referred to \citet{2001ApJ...563..172D}.

For the \pbar\ data, we consider the AMS~98 \citep{2002PhR...366..331A}
experiment on the shuttle, the balloon-borne experiments
IMAX~92 \citep{1996PhRvL..76.3057M}, CAPRICE~94 \citep{1997ApJ...487..415B},
WIZARD-MASS~91 \citep{1999ICRC....3...77B}, CAPRICE~98 \citep{2001ApJ...561..787B},
and the series of BESS balloon flights BESS~95+97 \citep{2000PhRvL..84.1078O},
BESS~98 \citep{2001APh....16..121M}, BESS~99 and 2000 \citep{2002PhRvL..88e1101A},
and BESS~2002 \citep{2005ICRC....3...13H}. We also add the
BESS Polar results \citep{2008PhLB..670..103B}. For BESS data,
we use the solar modulation level as provided in \citet{2007APh....28..154S},
based on proton and helium data.

	\section{Illustration of MCMC chains and PDF found with the three
	trial functions $q(\vectheta_{\rm trial}, \vectheta_{i})$}
  \label{s:illustration}

To compare the three trial functions (Sect.~\ref{s:trial_functions}),
a simple setup is retained: the B/C ratio observed from HEAO-3 data 
is used to constrain the model parameters $\{\vectheta_i\}_{i=1,\ldots, 3}= \{\lambda_0,\, R_0,\, \delta\}$,
i.e. Model~I.

\begin{figure}[!t]
\centering
\includegraphics[width=0.5\textwidth]{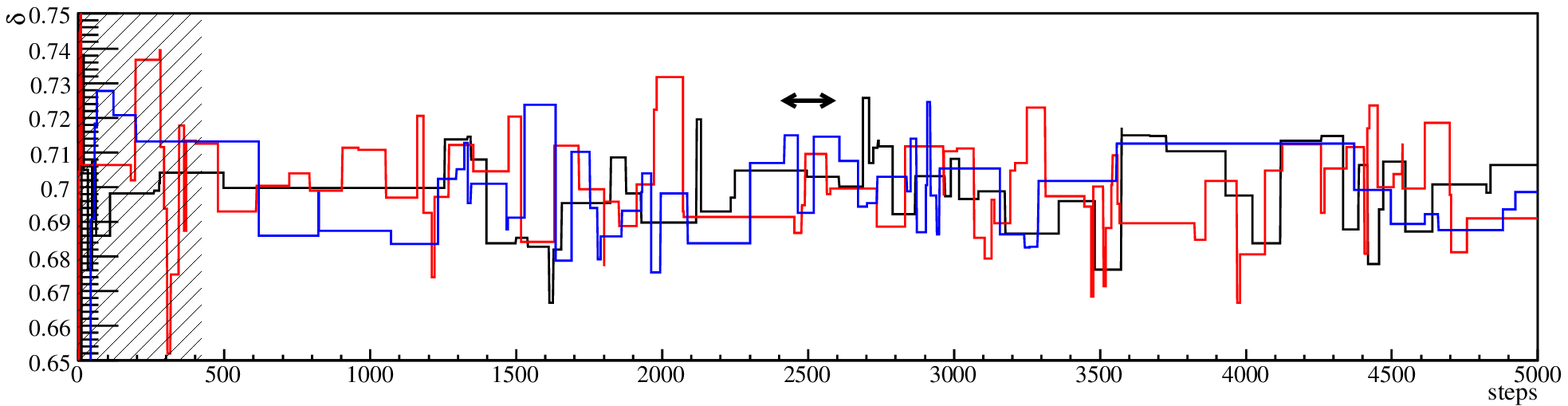}
\includegraphics[width=0.5\textwidth]{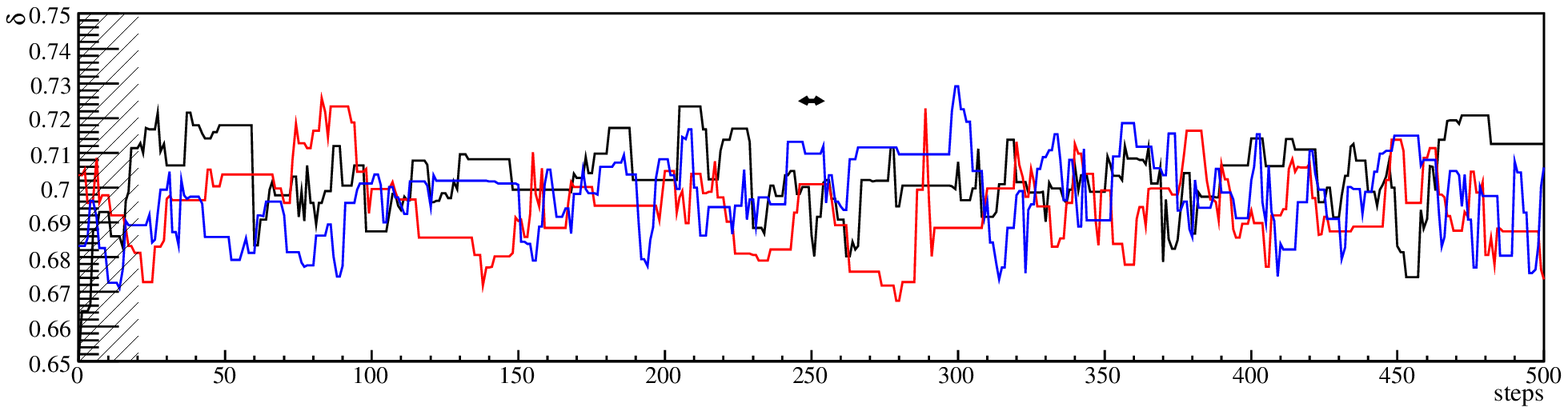}
\includegraphics[width=0.5\textwidth]{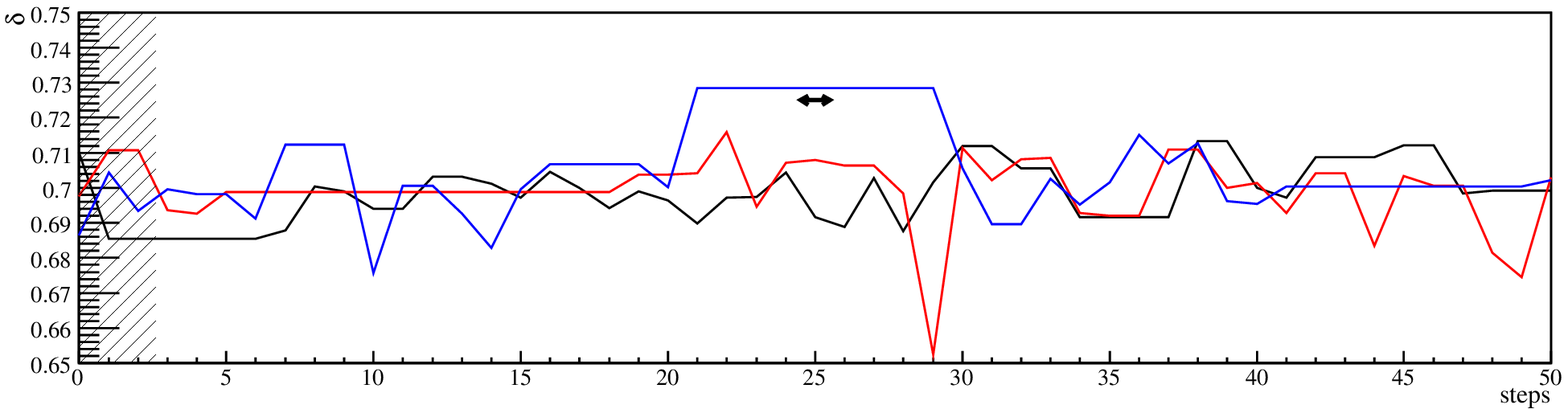}
\caption{Illustration of MCMC chains (here for the parameter $\delta$ and Model~I). From
top to bottom, chains generated from the Gaussian step, covariance matrix, and binary-space partitioning.
Three chains are shown in each panel: the shaded area corresponds
to the burn-in length and the arrow to the size of the correlation length $l$ defined by Eq.~(\ref{eq:correl_length}).
Although each process consists of 10~000 steps, the Gaussian step zoom in on the first 5~000 steps,
the two others displaying respectively 500 and 50 steps. This
indicates that each method allows a gain of $\sim 10$ in efficiency to extract the PDF
(compare the size of the arrows with the number of steps in each case).
}
\label{fig:chain}
\end{figure}
Taking advantage of parallel processing (as underlined in Sect.~\ref{s:implem}),
we combine several chains of 10~000 steps for each trial function. We start 
with the Gaussian trial function. It combines $N_c=40$ chains and
its output is used to calculate the covariance matrix. Taking
advantage of smaller burn-in and correlation lengths, only 20 chains
need to be combined for the covariance trial function. Again, the resulting
PDFs are used as input to the BSP trial function, which also combines
20 chains. Several chains (shown here for $\delta$), along with their burn-in
length and correlation step $l$, are shown in Fig.~\ref{fig:chain}.

\begin{figure*}[!t]
\centering
\includegraphics[width=0.47\textwidth,height=0.41\textwidth,keepaspectratio=false]{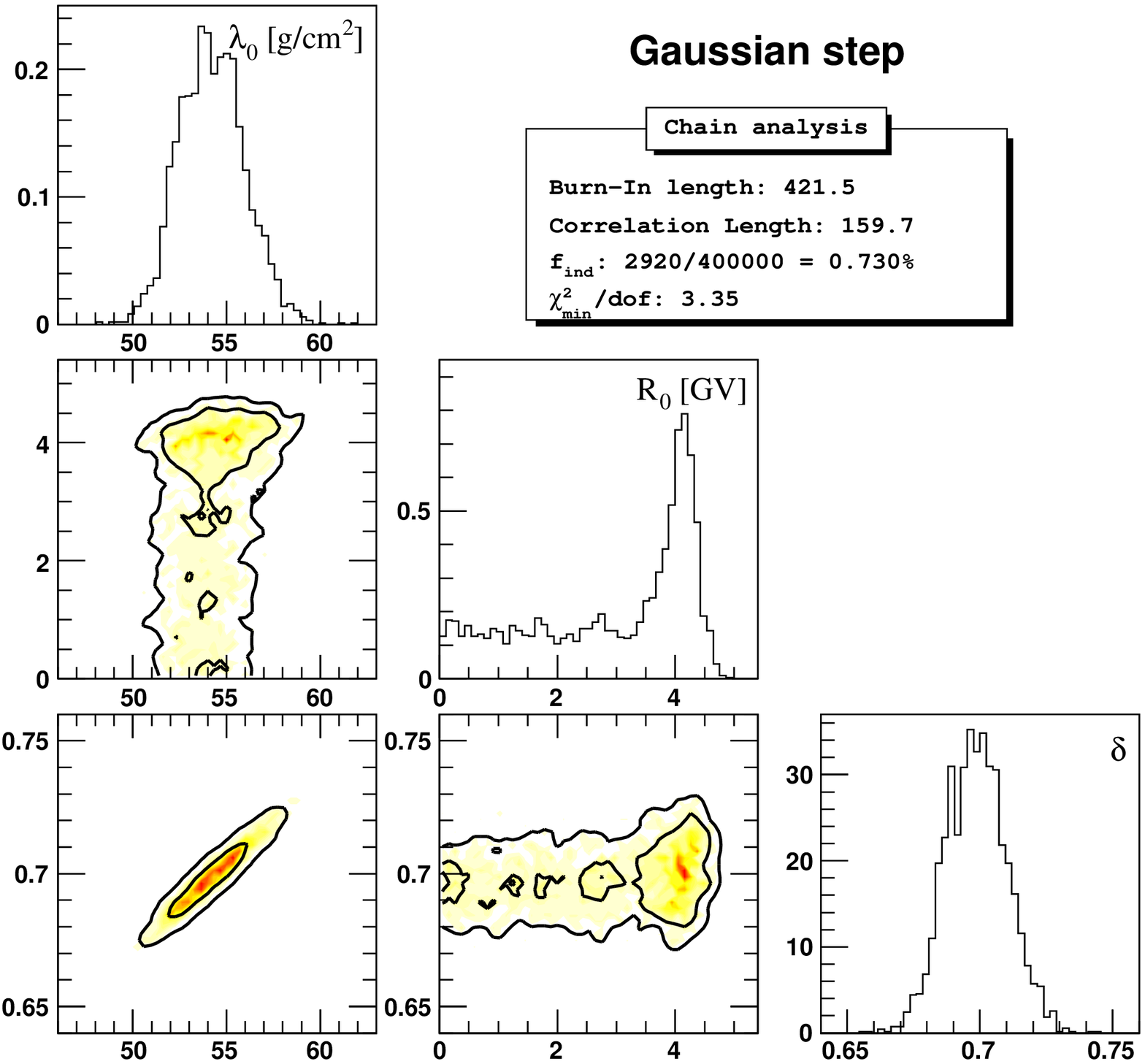}
\hspace{0.8cm}
\includegraphics[width=0.47\textwidth, height=0.41\textwidth,keepaspectratio=false]{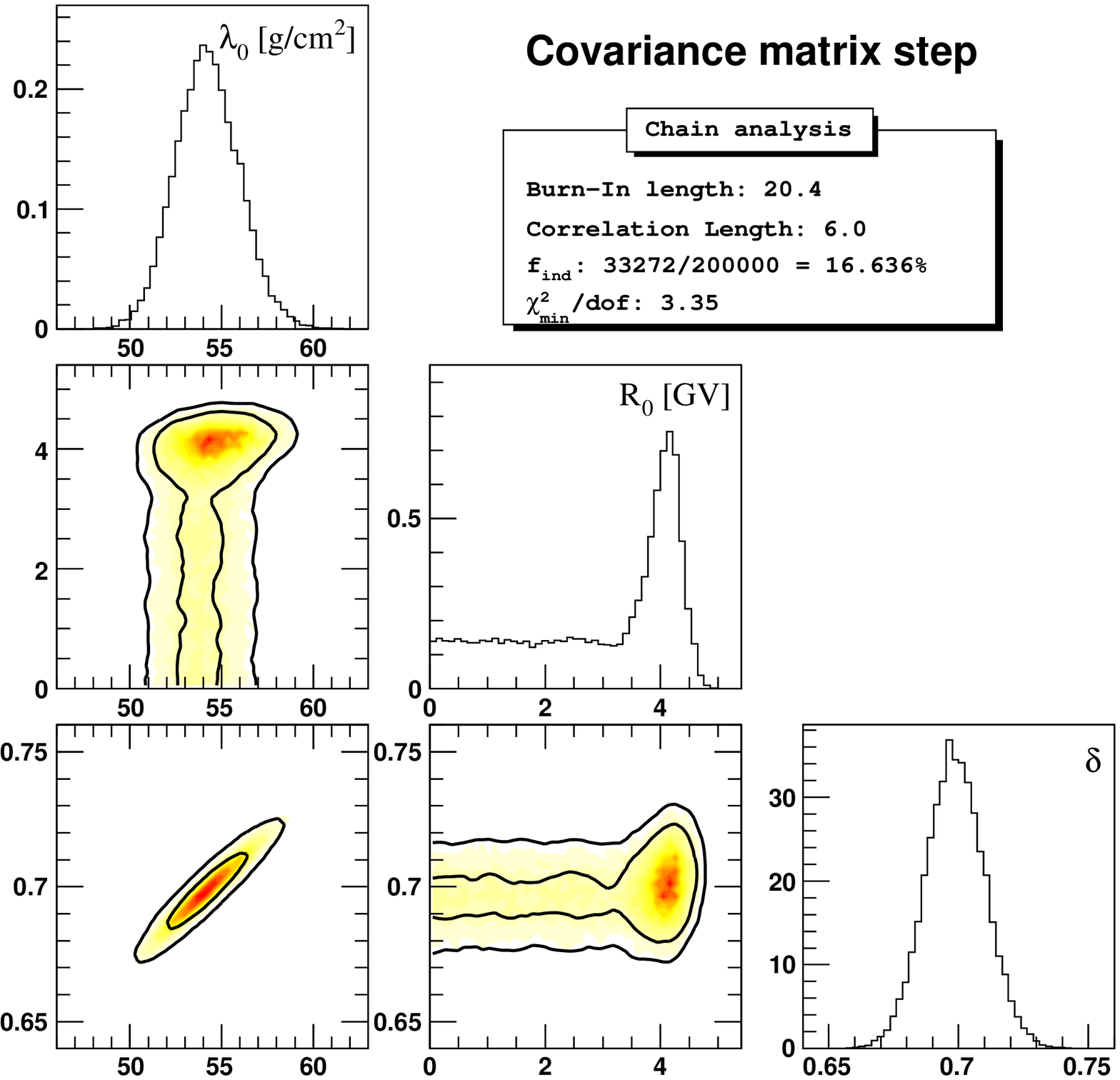}\vspace{0.6cm}
\includegraphics[width=0.47\textwidth,height=0.41\textwidth,keepaspectratio=false]{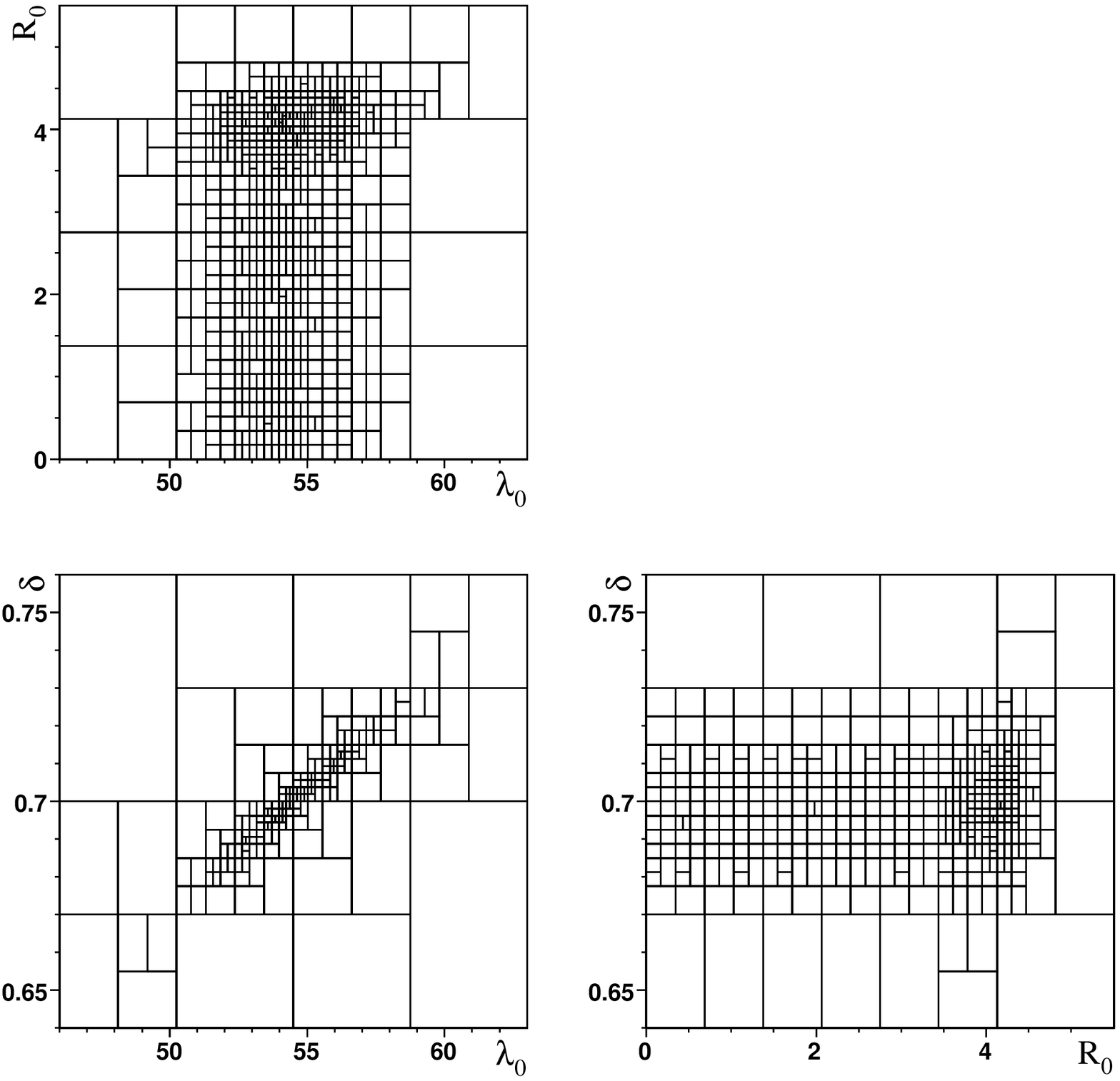}
\hspace{0.8cm}
\includegraphics[width=0.47\textwidth,height=0.41\textwidth,keepaspectratio=false]{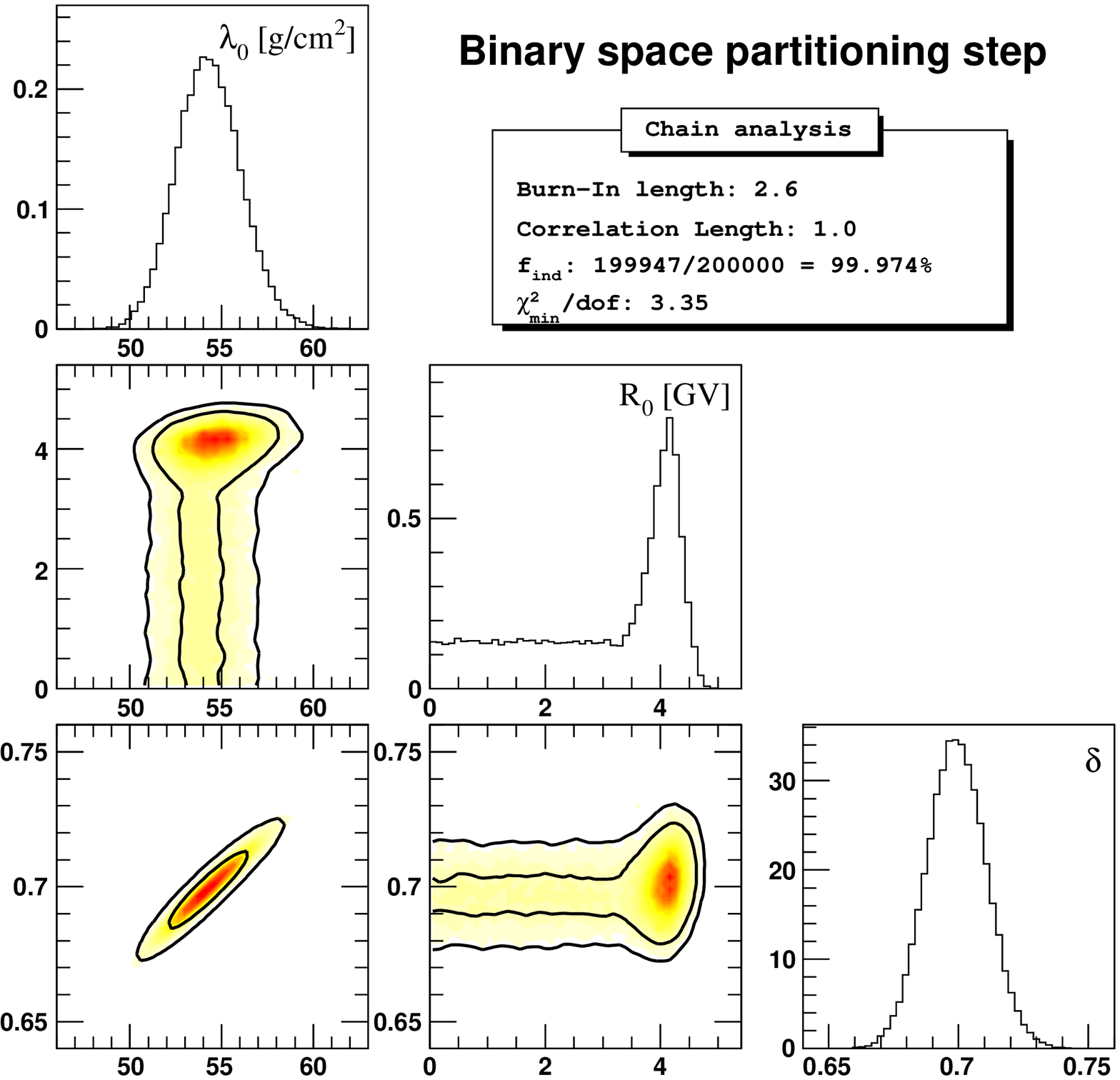}
\hspace{0.8cm}
\caption{Posterior PDF of $\{\vectheta^{(\alpha)}\}_{\alpha=1, \ldots, 3}= \{\lambda_0,\, R_0, \delta\}$ using the 3
proposal densities defined in Sect.~\ref{s:trial_functions} (Gaussian|top left;  Covariance matrix|top right; BSP|bottom right).
In these three panels, the diagonal shows the 1D marginalised posterior density function of the
indicated parameter. The number of entries in the normalised histograms corresponds to the number of uncorrelated steps
spent by the chain for values of $\lambda_0$, $R_0$ and $\delta$. Off-diagonal plots show the 2D marginalised posterior
density functions for the parameters in the same column and same line respectively: $\lambda_0-R_0$,
$\lambda_0-\delta$ and $R_0-\delta$. The colour code corresponds to the regions of increasing probability
(from paler to darker shade). The two contours (smoothed) delimit regions containing respectively 68\% and
95\% (inner and outer contour) of the PDF. Finally, the bottom-left panel shows an example of a 3D
parameter mesh (3 slices along the three different planes $\lambda_0-\delta-R_0$) obtained from the BSP method.}
\label{fig:B2CwoVa}
\end{figure*}

The result of the three sampling methods (Gaussian, covariance matrix, and BSP) for the PDF
are shown in Fig.~\ref{fig:B2CwoVa}.
The insert in each panel provides mean values (over the number of chains processed $N_c$)
for relevant parameters of the chain analysis (Sect.~\ref{sec:chain_analysis}): a decrease
in the burn-in length $b$ (421.5, 20.4 and 2.6) and the correlation length $l$ (159.7, 6 and 1)
is found when moving from the Gaussian to the BSP sampling method. The fraction
of independent samples [as defined in Eqs.~(\ref{f_ind1}) and (\ref{f_ind2})]
is $f_{\rm ind}=0.7\%$ for the Gaussian step, while nearly every step is valid and
uncorrelated (total of 99.9\%) for the BSP mode.
This confirms that, for a given number of steps, refined trial functions are more efficient
in extracting the PDF (note however that some improvement comes from the fact that each
method takes advantage of the previous step of calculation).

Figure~\ref{fig:B2CwoVa} (bottom-left) illustrates the binary-space partitioning
discussed in Sect.~\ref{s:trial_functions}. It
shows the projections of box sides on  
the three 2D planes $\lambda_0-R_0$, $\lambda_0-\delta$ and $R_0-\delta$ of the 
parameter space. The partitioning has been produced by the BSP method using the 
covariance matrix procedure presented on the same figure (top-right), where the box 
density is clearly proportional to the estimated target density.

\end{appendix}

\bibliographystyle{aa}
\bibliography{mcmcI}
\end{document}